\documentclass[aps,pre]{revtex4}
\usepackage{latexsym}
\usepackage{graphicx}
\usepackage{amsmath}
\usepackage{amssymb}
\usepackage{amsfonts}
\usepackage {color}

\usepackage[english]{babel}

\def \Sh{\text{Sh}}

\begin{document}

\title{Network models of dissolution of porous media}
\author{Agnieszka Budek}
\author{Piotr Szymczak}
\affiliation{Institute of Theoretical Physics, Faculty of Physics, University of Warsaw, Hoza 69, 00-618, Warsaw, Poland}

\begin{abstract}
We investigate the chemical dissolution of porous media using a network model
in which the system is represented as a series of interconnected pipes with the
diameter of each segment increasing in proportion to the local
reactant consumption. Moreover, the topology of the network is allowed
to change dynamically during  the simulation: as the diameters of the
eroding pores become comparable with the interpore distances, the
pores are joined together thus changing the interconnections within
the network. With this model, we investigate different growth regimes
in an evolving porous medium, identifying the mechanisms responsible for the emergence of specific patterns. We consider both the random and regular network and study the effect of the network geometry on the patterns. Finally, we consider  practically important problem of finding an optimum flow rate that gives a maximum increase in permeability for a given amount of reactant. 
\end{abstract}

\maketitle

\section{Introduction}

Chemical erosion of a porous medium is a complex process, involving an interplay between flow, transport, reaction and geometry evolution. The nonlinear couplings between these processes may lead to the formation of intricate dissolution patterns \cite{Daccord1987,Daccord1987a,Golfier2002b}, the characteristics of which depend strongly on the fluid flow and mineral dissolution rates. In particular, in a broad range of physical conditions, long, finger-like channels or ``wormholes'' are spontaneously formed, where the majority of the flow is focused. 

Understanding the details of a dissolution process is of fundamental importance in a variety of geological systems, including diagenesis, karst formation, aquifer evolution~\cite{Ortoleva1994}, and melt migration~\cite{Aharonov1995}. It also plays an important role in a number of engineering applications, such as dam stability~\cite{Romanov2003}, ${\rm CO}_2$ sequestration~\cite{Ennis-King2007}, risk assessment of contaminant migration in groundwater~\cite{Fryar1998}, and stimulation of petroleum reservoirs~\cite{Hoefner1988,Fredd1998}. 

Due to the complexity of the problem, analytical results are scarce and limited to either the initial stages of the process \cite{Chadam1986,Sherwood1987,Hinch1990,Szymczak2011a} or the analysis of simple model systems \cite{Nilson1990,Nagatani1991,Sahimi1987,Sahimi1991,Redner2001}. Thus most studies of the porous media dissolution rely on extensive numerical work. A variety of approaches has been followed in modelling, which can be classified into three major categories according to the length scales involved. The coarsest description is provided by the Darcy-scale models~\cite{Golfier2002b,Panga2005,Kalia2007} based on continuum equations with effective variables such as dispersion coefficients, Darcy velocity and bulk reactant concentrations. On the other side of the spectrum are pore-scale numerical simulations~\cite{Bekri1995,Kang2002,Kang2006} where the equations for fluid flow, reactant transport and chemical kinetics are solved in an explicitly three-dimensional pore space. Naturally, the pore-scale models are  much more accurate than the Darcy-scale ones but also highly expensive computationally, and thus limited by the system sizes that they can represent. Finally, somewhere in between these two levels of description are the  network models~\cite{Hoefner1988,Fredd1998,Daccord1989}, which model fluid flow and dissolution in a network of interconnected pipes, with the diameter of each network segment or pipe increased in proportion to the local reactant consumption. This is also the approach followed in the present work.

In constructing the model, we are largely following the ideas of Hoefner and Fogler \cite{Hoefner1988} but with a few important modifications. First, we take into account a potential limiting role of the  mass transfer of reactants to the pore surface. In that way, we obtain the description of the system in terms of two dimensionless parameters: the effective Damk\"{o}hler number, $\text{Da}_{eff}$, relating the reaction rate to the mean fluid velocity in the pores and another dimensionless number, $G$, measuring the extent to which the dissolution rate is hindered by diffusive transport of reactant across the aperture. 

Another important novel element that we introduce is to allow for dynamically changing topology of the network as the dissolution proceeds. As the diameters of the eroding pores become comparable to the interpore distances, the pores are joined together, thus changing the interconnections within the network. This allows for a more realistic representation of the evolving topography of the dissolving porous medium.

Although the original motivation for the construction of the Hoefner and Fogler network model \cite{Hoefner1988} was the analysis of acidization experiments in limestone cores, another natural application of the model is the simulation of early stages of karst formation in the limestone bedding planes which separate the individual strata in the rock. As elucidated by Ewers  \cite{Ewers1982} and Dreybrodt~\cite{Dreybrodt1988} in a first approximation one may regard a bedding plane as a two-dimensional porous medium with an average pore size comparable to the grain size of the confining rock, thus 2d network models seems to be particularly suited for simulations of such a system. 

The paper is organized as follows: In Sec.~\ref{model} we describe the network model used to represent the evolving porous medium. Results of the numerical simulations are given in Secs.~\ref{phase_diag}-\ref{breakthrough}. First, in Sec.~\ref{phase_diag} we analyze the form of the dissolution patterns as a function of flow velocity and surface reaction rate. In particular, we identify the regime in which the hierarchical structure of dissolution channels is formed and   study their length distribution. Then, in Sec.~\ref{regular} we consider the dissolution of regular lattices. Finally, in Sec.~\ref{breakthrough} we consider a problem of optimal injection rate, important in petroleum reservoir stimulation, so as to achieve the maximum increase in permeability for a given amount of reactant~\cite{Fredd1998,Golfier2002b,Panga2005,Kalia2007,Cohen2008a}.  We finish with a summary of our results and conclusions.

\section{Theoretical Model}\label{model}

The theoretical model used in this work is an extension of the Hoefner and Fogler scheme \cite{Hoefner1988}, who have pioneered the application of pore network models to study dissolution processes. Pore-network modelling has been extensively used in simulating single- and multiphase flow in porous media (see e.g. books by Dullien \cite{Dullien1992} and Sahimi \cite{Sahimi2011} for a detailed description of  different models and the discussion of their predictive abilities). However, the applications of pore models to dissolution processes remain scarce \cite{Hoefner1988,Daccord1989,Fredd1998,Li2006}. 

\begin{figure}[!h,przykladowa_siatka]
\centering
\includegraphics[angle=0,width=6cm]{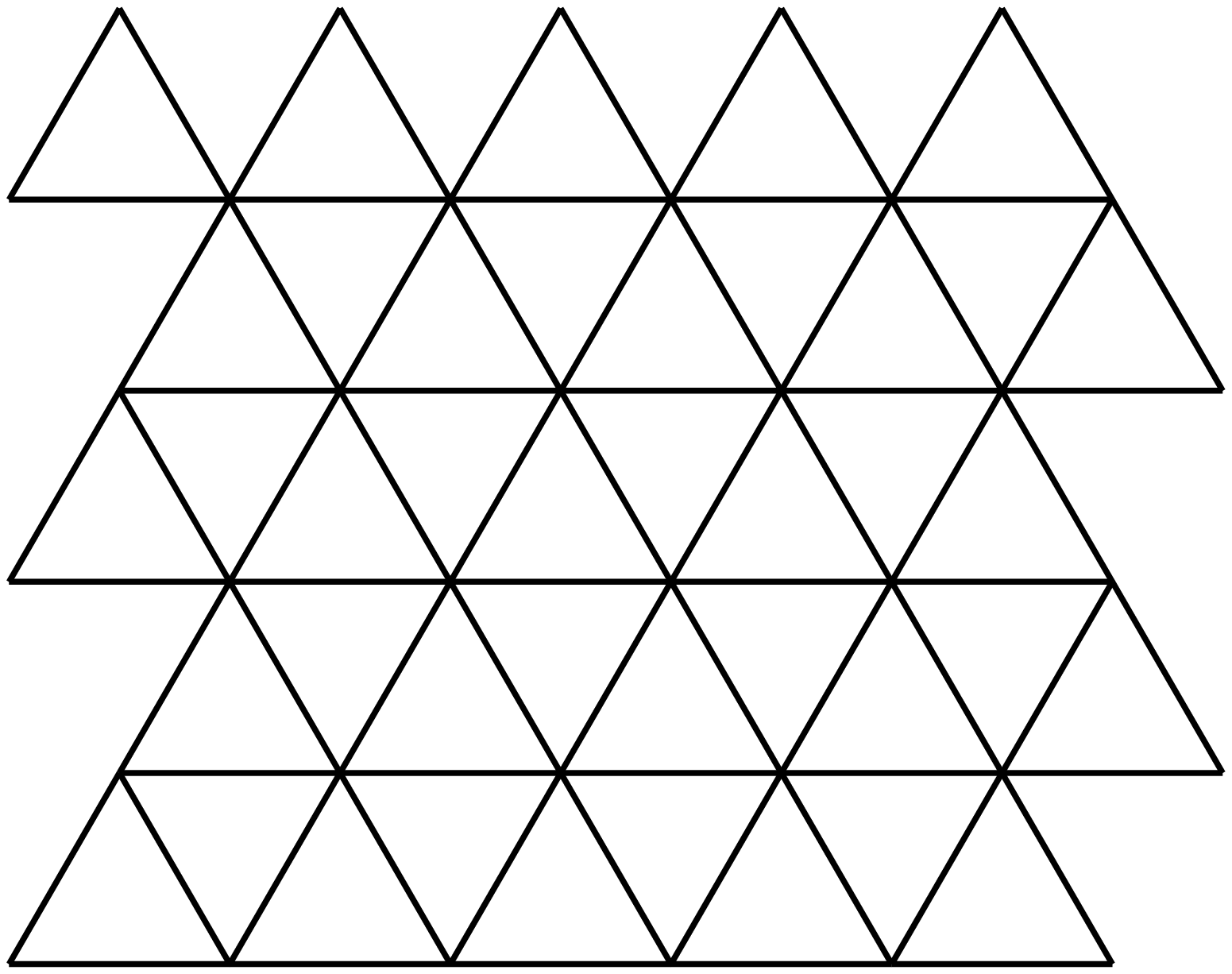}
\includegraphics[angle=0,width=6.5cm]{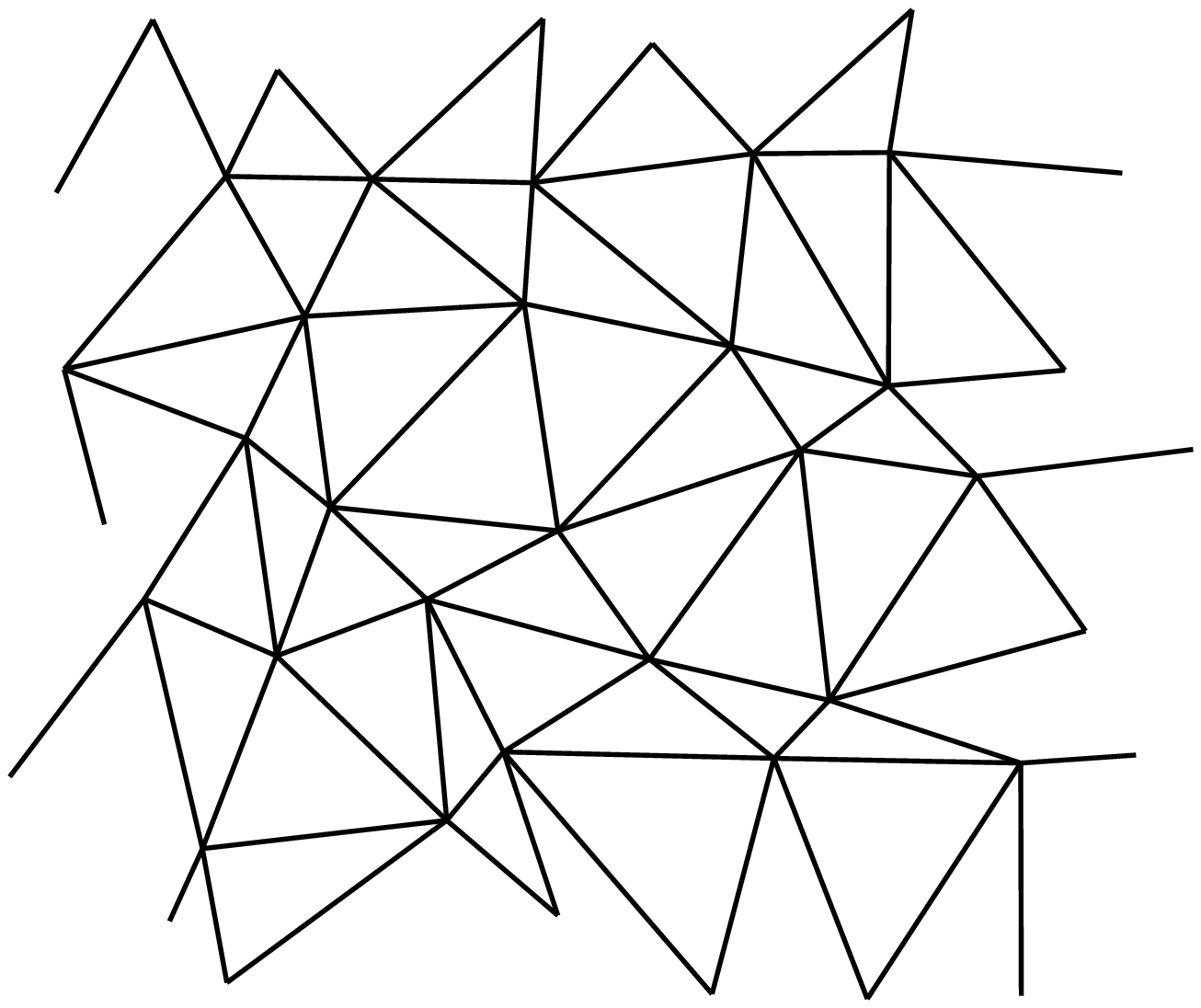}
\caption{
Triangular lattice with regular or random node configuration.}
\label{fig:przykladowa_siatka}
\end{figure}

In the model, the porous medium is represented as a triangular network of cylindrical tubes. The points where the tubes meet are referred to as nodes. The nodes can either be placed on a regular hexagonal lattice of lattice constant $l_0$ or randomly, as illustrated in Fig.~\ref{fig:przykladowa_siatka}. In the latter case, a random displacement, uniformly sampled from a value between $-0.4 l_0$ and $0.4 l_0$, is added to the  lattice nodes.

\begin{figure}[!h,kanalik_1]
\centering
\includegraphics[angle=0,width=8.5cm]{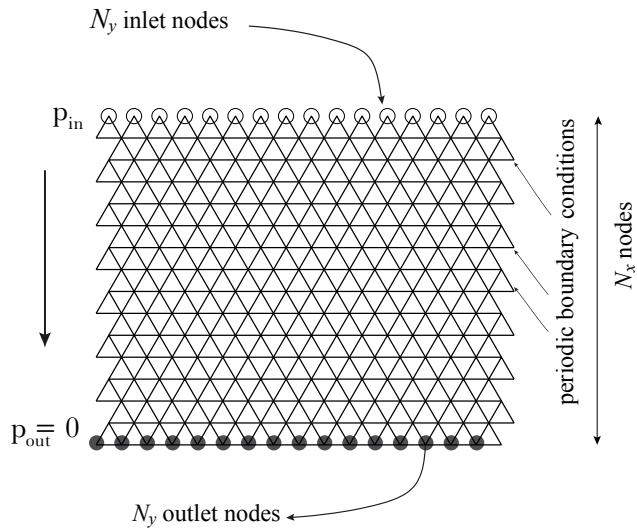}
\caption{
A schematic representation of a network with line inputs and outputs.}
\label{fig:schemat_po_calosci}
\end{figure}

Each pore is a tube with an initial diameter $d_0$, which gets enlarged during the dissolution process. 
A reactive fluid is injected into the network through a set of {\it inlet nodes}, where the pressure $p_{in}(t)$ is imposed, and leaves the system through {\it outlet nodes} where the pressure is kept at zero, $p_{out}=0$. Depending on the physical situation we intend to model, the inlet pressure is either kept constant, or it is being adjusted in each time step so as to keep the total flow through the system constant. The concentration of reactant in the incoming fluid is kept at a constant value of $c=c_{in}$. The inlet and outlet nodes can either be positioned at separate points within the network or placed along its two opposite boundaries,  as illustrated in Fig.~\ref{fig:schemat_po_calosci}. Additionally, periodic boundary conditions are applied along the lateral direction.

\subsection{Flow problem}

Fluid flow in the pores is described by the Hagen-Poiseuille equation for the volumetric flux 
 \begin{equation} 
q_{ij} = -  \frac{\pi} {128 \mu l_{ij}} d_{ij}^{4} (P_j-P_i),
\label{eq:Poiseuille}
\end{equation}
where $(P_j-P_i)$ denotes pressure drop along the pore joining node $i$ with node $j$, $q_{ij}$ is the volumetric flux in this pore, $d_{ij}$ and $l_{ij}$ are its diameter and length respectively  and  $\mu$ is the dynamic viscosity of the fluid. At each node, we also have a continuity condition
 \begin{equation} 
\sum_i q_{ij} = 0,
\label{eq:zachowanie}
\end{equation}
where the sum is over all the nodes connected by a pore with node $j$.  The resulting system of sparse linear equations can then be solved for pressure values at the nodes. 

\subsection{Dissolution of a single pore}

\begin{figure*}[!b,przykladowa_siatka]
\centering
\includegraphics[angle=0,height=3.5cm]{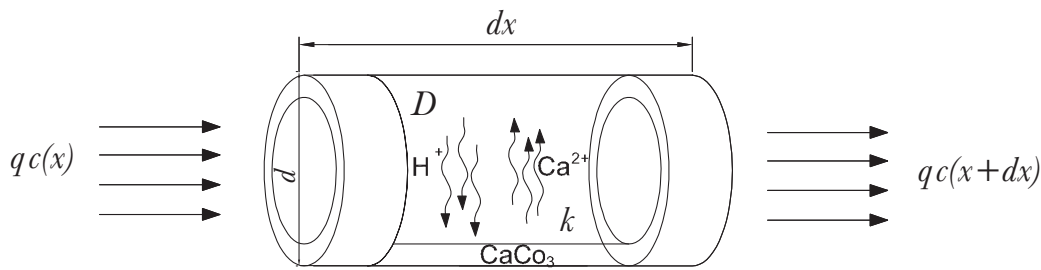}
\caption{
A schematic view of a pore and reactions at its surface.}
\label{fig:schemat_modelu}
\end{figure*}

Let us now consider one of the pores comprising the network and study the evolution of its diameter as a function of time. We assume that the walls of the pore are dissolving according to the linear rate law
\begin{equation}
J_r = k c_w.
\end{equation}
Here $J_r$ is the reactive flux (the number of absorbed reactant molecules per unit area and unit time), $k$ is the surface reaction rate and $c_w$ - the reactant concentration at the wall. The precise meaning of these variables depends on the particular reaction of interest, for example for carbonate dissolution by a strong acid, the respective reaction is 
\begin{equation} 
\textrm{CaCO}_3 + \textrm{H}^{+} \rightarrow \textrm{Ca}^{2+} + \textrm{HCO}_3^{-}. 
\label{eq:reakcja}
\end{equation}
In this case $c$ is a concentration of $H^{+}$ ions and $J_r$ describes the flux of hydrogen ions as illustrated in Fig. \ref{fig:schemat_modelu}. However, when calcite is dissolved by aqueous $\text{CO}_2$ at pH values similar to those of natural groundwater, the relevant variable is rather the calcium ion undersaturation \cite{Plummer1978,Dreybrodt1990}.
Note that in general the full chemistry of the carbonate dissolution is more complex \cite{Plummer1978}, but a single reactant description is a reasonable approximation in a broad range of chemical conditions \cite{Fredd1998,Dreybrodt1990}.

Additionally, we are taking into account the fact that, in order to take part in the reaction, the reactant first needs to diffuse from the bulk towards the pore surface. The diffusive flux can be expressed in terms of the difference between the surface concentration, $c_w$, and the bulk concentration, $c$ by using a mass-transfer coefficient or Sherwood number~\cite{Gupta2001},
\begin{equation}
\label{eq:RdiffSh}
J_{diff} = \frac{D \text{Sh}}{d}(c-c_w).
\end{equation} 
Here, the fluid bulk mean concentration of the reactant, $c$, also known as the mixing cup concentration is defined in terms of flow-weighted average of the concentration field over the cross-section of the pore:
\begin{equation}
c=\frac{1}{q}\int_0^{\frac{d}{2}} c(r) v(r) 2\pi rdr.
\end{equation}
In general, the Sherwood number, $\text{Sh}$, depends in a complicated way on reaction rate at the pore surface ($k$) but the variation is relatively small~\cite{Hayes1994,Gupta2001}, bounded by two asymptotic limits: high reaction rates (transport limit), $\text{Sh}=4.364$, and low reaction rates (reaction limit), $\Sh=3.656$. In the numerical calculations we approximate $\Sh$ by a constant value $\Sh=4$. Additionally, we neglect entrance effects, which otherwise make the Sherwood number dependent on the distance from the inlet of the pore, $x$.

By equating the reactive and diffusive fluxes $J_r = J_{diff}$ we can relate the wall concentration $c_w$ to the bulk one \cite{Gupta2001},
\begin{equation}
c_w = \dfrac{c}{1+kd/D\text{Sh}}.
\end{equation}
which can then be used to express the reactive flux in terms of the bulk concentration in the pore 
\begin{equation}\label{eq:Rcm}
J_r = k_{eff}c,
\end{equation}
where the effective reaction rate is given by
\begin{equation}\label{eq:keff}
k_{eff} = \dfrac{k}{1+kd/D\text{Sh}},
\end{equation} 
As seen from above, the relative importance of reactive and diffusive effects is described by the ratio 
\begin{equation}\label{eq:G}
g(d)=kd/D\text{Sh}
\end{equation}
In sufficiently narrow pores the diffusion is fast and able to keep the concentration field almost uniform across the pore diameter. In such a case, characterized by $g \ll 1$, the dissolution becomes reaction limited and $k_{eff} \approx k$. In the other limiting case, for wide pores and/or fast reactions, the reaction rate becomes hindered by diffusive transport of reactant across the aperture. As $g \gg 1$, the dissolution can become entirely diffusion limited with $k_{eff} \approx D\Sh/d$.
  
Next, the concentration of the reactant along the pore can be obtained from the mass balance equation. Neglecting axial diffusion, this leads to
\begin{equation}
\label{eq:balance}
q \frac{dc}{dx} = - \pi d k_{eff} c,
\end{equation} 
where $x$ is the coordinate along the axis of the pore. For a pore of a constant diameter, this can be solved to yield
\begin{equation} 
c(x) = c_0 e^{-\frac{\pi d k_{eff} x}{q}},
\label{eq:stezenia}
\end{equation}
with $c_0$ denoting the concentration of reactant at the inlet of a pore.

Finally, the erosion rate of the pore surfaces is described by 
\begin{equation}
\label{eq:erosion}
 \partial_t (d/2) =  \frac{ k_{eff}}{\nu c_{sol}} c,
\end{equation}
where $c_{sol}$ is  the concentration of soluble material and $\nu$ accounts for the stoichiometry of the reaction. The total volume of a mineral dissolved  from the walls of a pore of diameter $d$ over time $\Delta t$ is then
\begin{equation}
\Delta V =  \frac{\pi d k_{eff} \Delta t}{\nu c_{sol}} \int_0^l c(x) dx = \Delta t q \frac{c_0}{c_{sol} \nu}  (1-e^{-\pi d k_{eff} l/q}).
\end{equation} 
In order to keep the model tractable, we assume that the pores are broadening uniformly along their length. Thus the above volume change corresponds to the enlargement of the pore diameter by
\begin{equation}
\Delta d =\frac{\Delta V}{\pi d l} = \frac{2 \Delta t q}{\pi d l} \frac{c_0}{c_{sol} \nu}  (1-e^{-\pi d k_{eff} l/q}).
\label{eq:dlugosci}
\end{equation} 
As seen from the above, the decay of the concentration along the length of the pore is determined by a function
\begin{equation}
f(d,q)=\frac{\pi d k_{eff} l}{q}=\frac{\pi d k l / q}{1+g(d)},
\label{eq:Da}
\end{equation} 
which relates the reaction rate to the rate of convective transport. Using $f$ and $g$ one can rewrite formulae \eqref{eq:dlugosci} and \eqref{eq:stezenia} in a simple form
\begin{equation} 
\frac{\Delta d}{d_0} = \frac{\Delta \hat{t} c_0/c_{in}}{(1+g)f} (1-e^{-f}),\qquad
\label{eq:d_zwiezle}
\end{equation}
and
\begin{equation} 
c(x)=c_0 e^{-f x/l}.
\label{eq:C_zwiezle}
\end{equation}
where the time has been scaled by
\begin{equation}
\hat{t} = 2  k \gamma t/d_0,
\label{timess}
\end{equation}
and 
\begin{equation}
\gamma = \frac{c_{in}}{\nu c_{sol}}
\label{acid}
\end{equation}
is the acid capacity number or volume of solid dissolved by a unit volume of reactant.

\subsection{Dimensionless parameters characterizing the evolution in the network}

The functions $f$ and $g$ characterizing the dissolution in the pores depend on time through $d$ and $q$. Additionally, their values can vary across the sample, since the pores have different lengths and diameters and carry different flows.  In the following, while characterizing the effects of flow rate and surface reaction rate on the dissolution patterns, we will construct the phase diagram of the patterns in terms of characteristic values of these parameters at $t=0$
\begin{equation} 
\text{Da}_{eff} = \frac{\pi d_0 k l_0 /q_{in}}{1+\frac{k d_0}{\text{Sh} D}}
\label{Daeff}
\end{equation}
and
\begin{equation}\label{eq:G0}
G=kd_0/D\text{Sh},
\end{equation}
where $l_0$ is the lattice constant of the underlying hexagonal network, $d_0$ is the initial pore diameter, whereas $q_{in}$ is a characteristic flow in the inlet pores, i.e. 
$$
q_{in}=Q/N_{inlet},
$$
where $N_{inlet}$ is the number of inlet pores. The first of these parameters, $\text{Da}_{eff}$, can be interpreted as an effective Damk\"{o}hler number relating the reaction rate to the mean fluid velocity in the pores, whereas $G$, as has already been mentioned, measures the extent to which the dissolution rate is hindered by diffusive transport of reactant across the aperture. The parameter $G$ plays a similar role to the Thiele modulus in chemical engineering \cite{Finlayson1980} which measures the relative importance of diffusion and reaction.

\subsection{Implementation of the model}

The above-described model is implemented numerically in the following way:
\begin{itemize}
\item Pressure in each node and flow through each tube are calculated from equations \eqref{eq:Poiseuille} and \eqref{eq:zachowanie} using MUltifrontal Massively Parallel Solver (MUMPS) \cite{MUMPS:1,MUMPS:2}.
\item The concentration field in each pore, starting from the inlet ones, is then obtained by a repetitive use of Eq.~\eqref{eq:C_zwiezle} under the assumption that at the pore intersections the flow is divided accordingly to the total volume flux through each tube    
\item Finally, the diameters of the pores are updated according to Eq.~\eqref{eq:dlugosci}.
\end{itemize}

\begin{figure}
\centering
\includegraphics[angle=0,height=4cm]{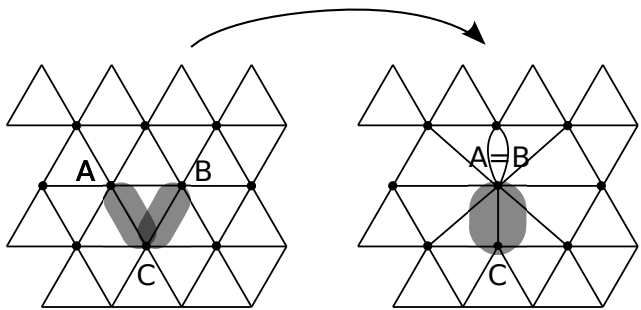}
\caption{Change of the network topology associated with pore merging. When the sum of the diameters of the pores AC and BC becomes larger than $2l_0$ the points $A$ and $B$ are joined and the two pores are replaced by a single pore with a diameter equal to $d_{AC}+d_{BC}$. 
}
\label{joining}
\end{figure}

Additionally, we introduce the possibility of merging the pores when their diameters  become comparable to the interpore distances.  To be more specific, when $d_i+d_j$ becomes larger than $2l_0$, we replace both pores by a single pore of diameter $d=d_i+d_j$, as illustrated in Fig.~\ref{joining}. Note that in this way we assure 
that the total reactive surface area before and after merging is the same, the condition which is crucial to guarantee that the overall reactant balance will not change as an effect of merging.  On the other hand, the total volume of the pores is not conserved during merging and the overall hydrodynamic resistance decreases. This, however, has only a minor effect on the system, since at the moment of merging the diameters of the pores are large and thus their resistances are almost negligible. 

An important parameter controlling the merging process is the ratio of the initial pore diameter $d_0$ to the lattice constant,  $l_0$. Since the condition for merging is $d_i+d_j \geq 2l_0$, in the systems with larger $d_0/l_0$ the pores will merge faster. In real rocks, the values of pore  aspect ratio span a rather broad range, depending on the type of porosity present in the rock structure \cite{Gueguen1994}: whether it is a network of interconnected microcracks ('crack porosity' with aspect ratios as low as $0.1 \%$) or intergrain pore space ('equant porosity' with aspect ratios of $0.1-1$).

\section{Characterization of dissolution patterns}\label{phase_diag}

\begin{figure*}[!b,stale_Q,onecolumn]
\centering
\includegraphics[angle=0,height=10cm]{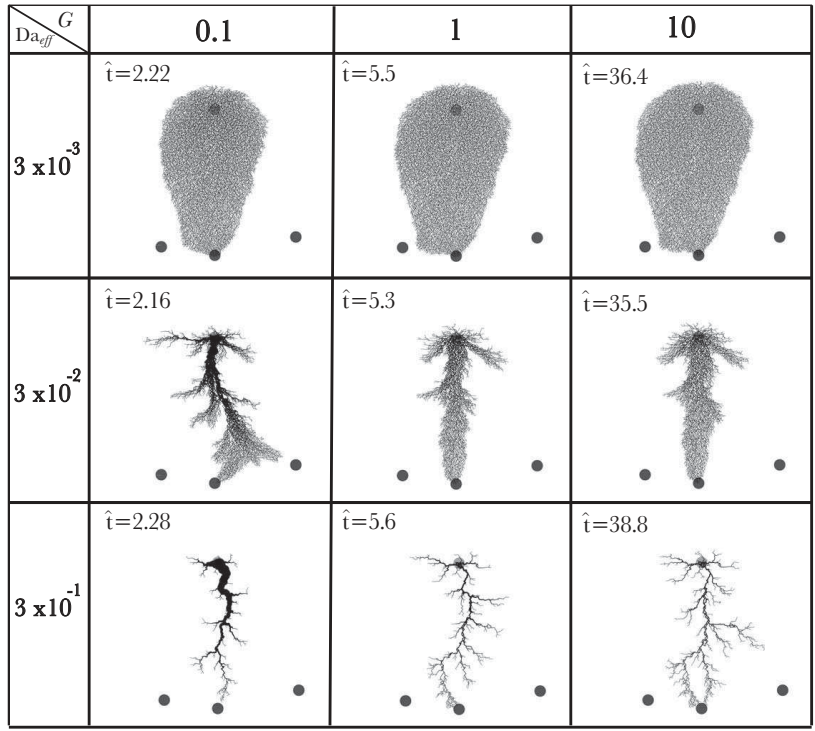}
\caption{
The patterns emerging in the dissolution of  $200\times 200$ lattice with one inlet and three outlets (marked by circles) in a range of different $\text{Da}_{eff}$ and $G$ numbers.  The effective Damk\"{o}hler number is the main factor responsible for the way the reactant is distributed throughout the system, whereas $G$ controls the amount of dissolution in the pores near the inlet. The simulations were performed under the fixed pressure drop conditions. The frames in the figure correspond to the breakthrough moment, and the  corresponding dimensionless time, $\hat{t} = 2  k \gamma t/d_0$, is shown for each frame.
}
\label{fig:Da_PeDa_stale_P}
\end{figure*}

Fig.~\ref{fig:Da_PeDa_stale_P} illustrates typical dissolution patterns for $200 \times 200$ random lattice over a range of different flow and reaction rates. 
There is one inflow in the system and three outflows situated at equal distances from the inlet, but not symmetrically. The pores are initially uniform in diameters. The widths of the lines representing the pores in the figure are proportional to their diameters. For the sake of clarity, we plot only the pores which have reached $d=3 d_0$ at a given time. The frames in the figure correspond to the 'breakthrough' moment when the dissolution reaches the outlet of the system, that is at least one of the outlet pores have broaden 3 times. Unless otherwise stated the value of pore aspect ratio was taken to be $d_0/l_0=0.025$. 

The phase diagram is plotted in terms of dimensionless parameters $\text{Da}_{eff}$ and $G$, defined in the preceding section.  For small $\text{Da}_{eff}$, due to the low reaction rate and large flow,  the reagent spreads evenly throughout many parallel pores along the main flow path and erodes them almost uniformly, which results in a diffuse, turnip-shaped pattern. This can be rationalized by noting that the characteristic penetration length of the reactant is, according to Eq. \eqref{eq:C_zwiezle}
\begin{equation}
l_p = l_0/\text{Da}_{eff} 
\label{penetr}
\end{equation}
thus for $\text{Da}_{eff} \approx 10^{-3} - 10^{-2}$ most of the pores along the main flow path will be invaded almost instantaneously.

\begin{figure}
\centering
\includegraphics[angle=0,width=7cm]{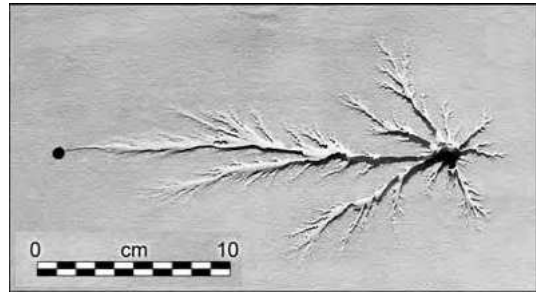}
\caption{
Dissolution patterns obtained by Ewers in the experiments on the dissolution of blocks of Paris plaster \cite{Ewers1982}. The inlet and outlet of the fluid are marked by circles.}
\label{Ewers}
\end{figure}

 However, as $\text{Da}_{eff}$ is increased, the penetration length is reduced and only the pores in the close vicinity of the inlet are invaded at the beginning of the dissolution process. The reagent is then consumed very quickly and further advancement of the dissolution front is only possible when one of the pores increases its flow rate at the expense of its neighbors. Such a competition between different flow paths is a characteristic feature of the dissolution problems \cite{Chadam1986,Daccord1987a,Daccord1987,Golfier2002b,Szymczak2006,Szymczak2009}: a path which gets a slightly larger flow dissolves faster than its neighbors, which decreases its hydraulic resistance and makes the flow there even higher. As a result, an intricate, fractal-like structure of channels is formed, where all the flow and dissolution is focused. The structure becomes increasingly more branched as $\text{Da}_{eff}$ is increased, with individual branches again competing with each other for the available flow, which leads to the final appearance of a single flowpath joining the inlet and outlet.

\begin{figure*}[!b,po_calosci_stale_P_raster]
\centering
\includegraphics[angle=0,height=15cm]{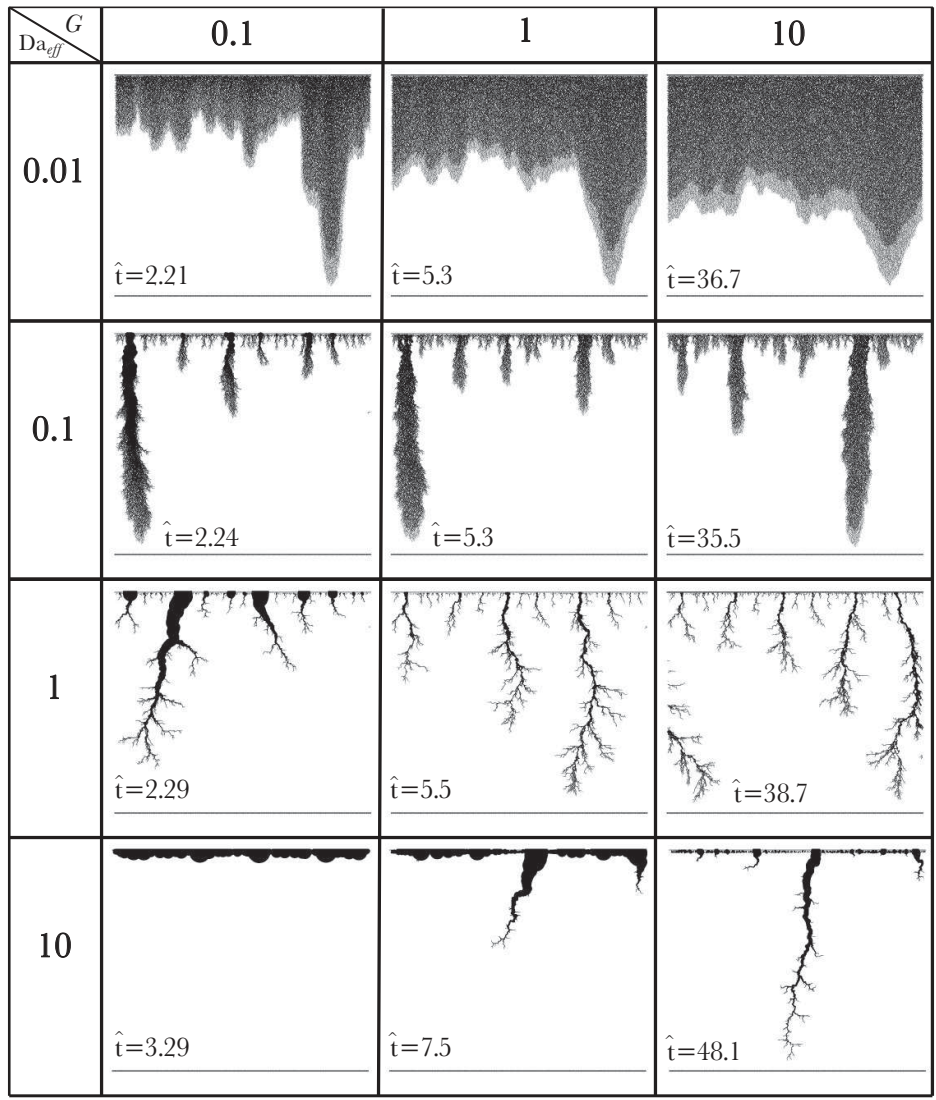}
\caption{
Phase diagram of dissolution patterns for different $\text{Da}_{eff}$ and $G$ in the system with line inlets and outlets. The simulations were performed under the fixed flow conditions. The frames in the figure correspond to either the breakthrough moment or to the moment when one of the channels has increased its initial diameter 300 times (the frames in a lower left corner). The corresponding values of the dimensionless time \eqref{timess} are shown for each frame.
}
\label{fig6}
\end{figure*}

The $G$ parameter affects the dissolution process in a more subtle way, much less evident compared to the effects caused by $\text{Da}_{eff}$. For larger values of $G$ we observe more branching in the wormholes and less broadening around the inlet, particularly at higher effective Damk\"{o}hler numbers.
This behaviour is connected with a simple mechanism that leads to a drastic increase  of dissolution  rate in the widest pores in the reaction ($G \rightarrow 0$) limit - the wider the pore is, the larger its surface area ($S = \pi d l$) becomes, so that it consumes more reactant compared to other pores. This leads to a situation in which increasingly larger amount of reactant is consumed in the first few pores in a given flowpath, and the reactant withdraws from the more remote pores (see also a more detailed discussion of this effect in the Appendix). This effect is most pronounced when a constant flow is imposed on the system. In the constant pressure case, the flow rate increases during the dissolution, hence $\text{Da}_{eff}$ decreases and the reactant is penetrating deeper inside the system, not being spent in the first few pores.

This effect is strongly reduced at larger values of $G$, where the dissolution relatively quickly switches to a transport-limited mechanism, which tends to slow down the dissolution rate as the pore opens, due to the inverse dependence of $k_{eff}$ on the diameter, as observed in Eq.~\eqref{eq:keff}.

The geometry of Fig.~\ref{fig:Da_PeDa_stale_P}, with point inputs and outputs, is relevant to a number of karst formation problems \cite{Ewers1982,Dreybrodt1988,Palmer1991}. The point inputs then represents particular sinks through which the water may enter the bedding plane, whereas 
outputs correspond to the springs draining the area. Ewers  \cite{Ewers1982} has also modelled such a situation experimentally by creating an artificial model of a bedding plane between the soluble plaster block and a transparent insoluble lower boundary and flushing the system with water through a number of point inlets and outlets. An example result of such an experiment is presented in Fig.~\ref{Ewers} and bears a lot of resemblance with the patterns obtained by use of a network model

Another experimentally important geometry is the one in which the reagent enters 
through the entire top surface and leaves through the bottom.  This is the geometry relevant to the experiments and simulations by Hoefner, Fogler and Fredd \cite{Hoefner1988,Fredd1998} on the dissolution of limestone blocks and also Golfier \cite{Golfier2002b} on the salt dissolution in a Hele-Shaw cell. In this geometry there are no preferred flow paths along which water may flow more rapidly than through others, thus initially many independent wormholes are formed which then compete for the available flow. 

Results of numerical simulation for different values of $\text{Da}_{eff}$ and $G$ for this geometry are presented in Figure \ref{fig6}. 
Constant total flow through the system has been imposed in these simulations (the development of the patterns in the analogous simulations under constant pressure conditions is represented in the movies in the Supplementary Material \cite{supp}). 
Even though an identical initial geometry has been used in each of the runs, one 
observes that position of the main wormhole varies for different $\text{Da}_{eff}$ and $G$, illustrating sensitivity of the dissolution process to physical parameters beyond the geometry.        

\begin{figure}
\centering
\includegraphics[angle=0,width=8.5cm]{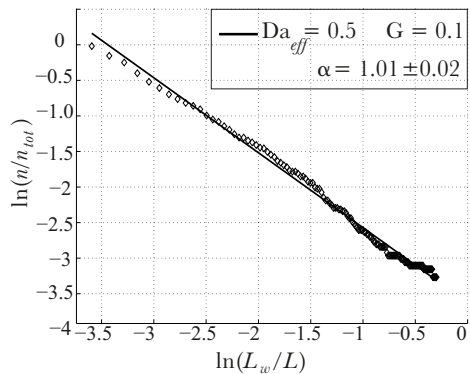}
\caption{
The distribution of wormhole lengths, $n(L_w)$, at the breakthrough time for $\text{Da}_{eff}= 0.5$ and $G =0.1 $. The solid line corresponds to the fit $n(L_w)\backsim L_w^{-\alpha}$. Wormholes shorter than $0.03 L$ and longer than $0.75 L$ (with $L$ the system length) were not taken into account in the distribution. The lengths are normalized by the system length and $n$ is normalized by the total number of wormholes in the sample, $n_{tot}$. Results correspond to a simulation with fixed total flow through the system.}
\label{fig:statystyka_dopasowanie_Q}
\end{figure}

The overall dependence of the dissolution patterns on $\text{Da}_{eff}$ and $G$ shares many similarities with that observed for point inlets and outlets. Again, the effective Damk\"{o}hler number is the key parameter controlling the penetration length - for small values of $\text{Da}_{eff}$, as the penetration length becomes comparable with the size of the system, dissolution is uniform. Then, as $\text{Da}_{eff}$ is increased, a well-defined reaction front appears, which becomes unstable in the course of evolution. The linear stability analysis shows that the wavelength of this instability decreases with increasing Damk\"{o}hler number  \cite{Szymczak2011a}, which is  consistent with the results of present simulations - for $\text{Da}_{eff}=10^{-2}$ one observes long-wavelength undulations of the dissolution front whereas for larger $\text{Da}_{eff}$ separated wormholes are formed where the majority of the flow and reaction is focused, while most of the pore space is bypassed. To further investigate the scaling of the instability wavelength with the Damk\"{o}hler number we conducted a series of dissolution simulations of regular lattices with uniform pore lengths $l_0$ but with a very small random noise in the initial pore diameter distribution: $\Delta d/d_0 = 10^{-4}$. The results show sinusoidal modes developing in the dissolution front, as illustrated in Fig.~\ref{waves}. The wavelength is inversely proportional to the Damk\"{o}hler number, confirming that penetration
length, $l_p=l_0/\text{Da}_{eff}$, is the only important length scale in the early stages of dissolution, which is a common feature of the convection-dominated dissolution processes \cite{Szymczak2011,Szymczak2011a}.

The wormholes again compete in a process in which the longer wormholes drain flow from the shorter ones, limiting their growth. This time however, in contrast to the point inlet case, due to the translational invariance of the system the process of wormhole competition becomes self-similar.  The characteristic length between active (growing) wormholes increases with time, while
the number of active wormholes decreases, which leads to a scale-invariant, power-law distribution of wormhole lengths,
\begin{equation} 
n(L_w)\backsim L_w^{-\alpha},
\label{eq:prawo_potegowe}
\end{equation}
were $n(L_w)$ denotes number of wormholes longer than $L_w$. The fits to Eq.~\eqref{eq:prawo_potegowe} were performed in the range $0.5 \geq \text{Da}_{eff} \geq 5$ where the wormholes are well-pronounced and there are relatively many of them, which allows for the  adequate statistics. Wormholes shorter than $0.03 L$ and longer than $0.75 L$ (with $L$ the system size) were not taken into account in the distribution: the former because their lengths are influenced by the lattice discretization effects, the latter - because the longest wormholes remain active and the selection process there has not yet been concluded.

The values of the exponent $\alpha$ obtained from the fitting procedure over a range of different $\text{Da}_{eff}$ and $G$ numbers are presented in Table 
I. The exponent vary slightly with $G$ with the largest values corresponding to the reaction limited case (small $G$). 
An example distribution  of the wormhole lengths obtained at $\text{Da}_{eff}=0.5$ and $G=0.1$ is presented in Fig.~\ref{fig:statystyka_dopasowanie_Q} together with the respective fit. The reported values of $\alpha$ in reaction-limited regime (small $G$) are in a good agreement with the 2d Laplacian needle growth model of wormhole-wormhole competition \cite{Krug1993}, which yields $\alpha=1$.

\begin{table}
$\begin{array}{|c|ccc|}
\hline  
\text{Da}_{eff} \backslash G  & 0.1 & 1 & 10 \\ \hline  
 0.5 & 1.01 \pm 0.02 \hspace{0.1cm}&\hspace{0.1cm} 0.98 \pm 0.03 \hspace{0.1cm}&\hspace{0.1cm} 0.96 \pm 0.05\\ %\hline  
 1   & 0.98 \pm 0.03 \hspace{0.1cm}&\hspace{0.1cm} 0.90 \pm 0.02 \hspace{0.1cm}&\hspace{0.1cm} 0.89 \pm 0.02\\ %\hline  
 2   & 1.01 \pm 0.02 \hspace{0.1cm}&\hspace{0.1cm} 0.91 \pm 0.03 \hspace{0.1cm}&\hspace{0.1cm} 0.88 \pm 0.02\\\hline 
\end{array}
$
\label{tab:ws_alfa}

\caption{Parameter $\alpha$ for different $\text{Da}_{eff}$ and $G$.}
\end{table}

\begin{figure*}[!b,po_calosci_stale_P_raster]
\centering
\includegraphics[angle=0,height=15cm]{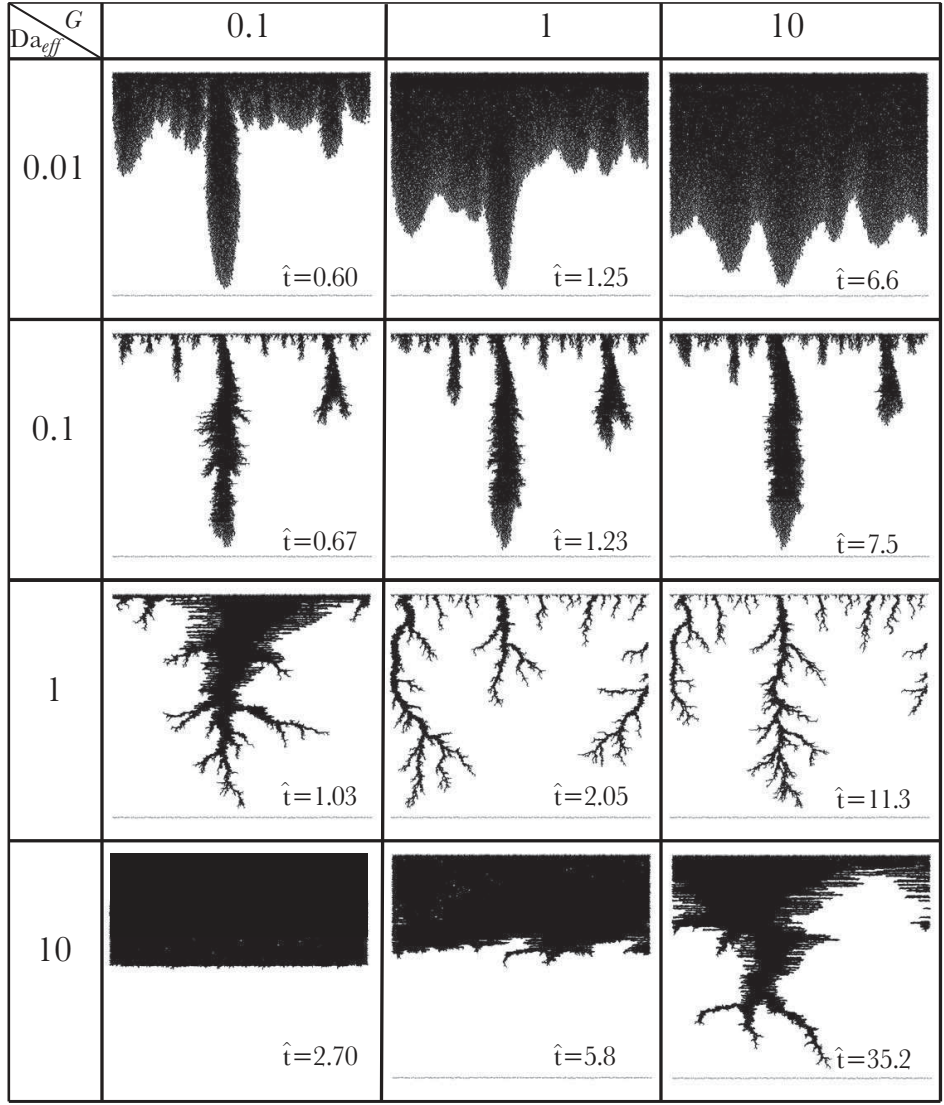}
\caption{
Same as in Fig.~\ref{fig6} but with pore merging (with pore aspect ratio $d_0/l_0=0.1$). The frames in the figure correspond to the breakthrough moment except for the two frames in the lower left corner, where, for the sake of legibility, we stop the simulation at the moment when half of the system has dissolved.
}
\label{fig8}
\end{figure*}

Finally, let us discuss the influence of pore merging on the dissolution patterns. The data in Figs.~\ref{fig:Da_PeDa_stale_P} and \ref{fig6} have been obtained in the simulation which has not allowed for the pore merging. For comparison, Fig.~\ref{fig8} shows the patterns obtained using pore merging simulation (with the initial pore aspect ratio  $d_0/l_0=0.1$). Comparing Figs.~\ref{fig6} and \ref{fig8} one notes that merging affects mainly the patterns corresponding to the large values of $\text{Da}_{eff}$. For small $\text{Da}_{eff}$ the dissolution is uniform, and  merging takes place mostly behind the scalloped dissolution front, hence it is not affecting the front instability dynamics. Then, as $\text{Da}_{eff}$ increases, the merging intensifies, in particular near the inlet and along the wormholes. Finally, for large $\text{Da}_{eff}$ and small $G$ the non-merging and merging patterns are contrastingly different - whereas in the non-merging case dissolution concentrates at the inlet, merging allows the matrix at the inlet to become completely dissolved and a steadily advancing front appears, separating fully dissolved pore space from the undissolved one. 

\begin{figure}[!h,przykladowa_siatka]
\centering
\includegraphics[angle=0,width=14cm]{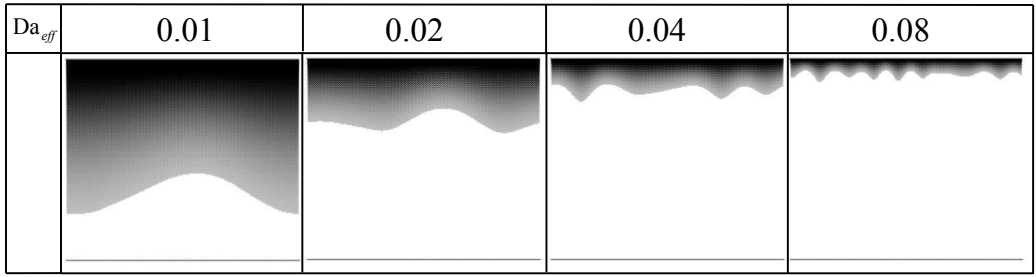}
\caption{
Concentration profiles in a dissolving pore network with a small initial noise, $\delta d/d_0 = 10^{-4}$. The shades of gray represent the magnitude of the concentration field with black standing for $c=c_{in}$ and white for $c < 0.07 c_{in}$.}\label{waves}
\end{figure}

\section{Dissolution of the regular lattice}\label{regular}

\begin{figure}
\centering
\includegraphics[angle=0,width=10cm]{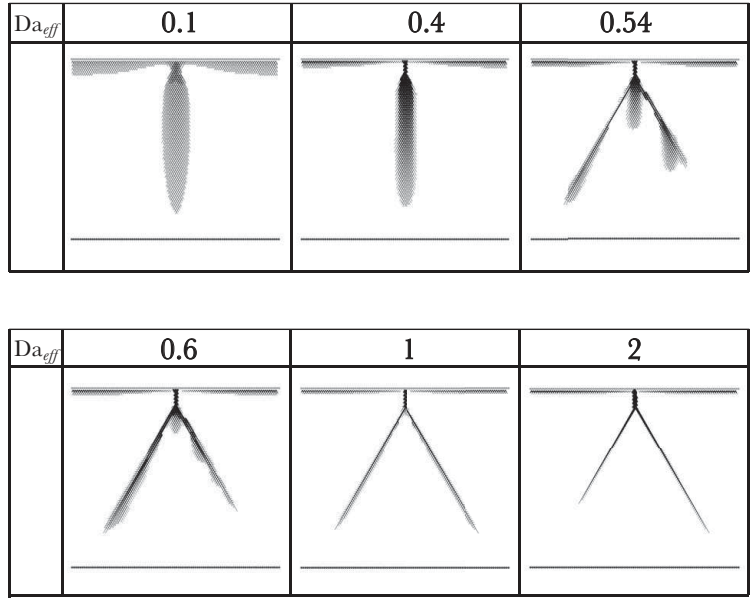}
\caption{
Dissolution patterns of a regular network for different $\text{Da}_{eff}$ and $G=1$. Initially, a short channel was cut in the center of inlet manifold to induce the localization of the flow. }
\label{fig:symmetric}
\end{figure}

In this section we consider the dissolution of an ideal hexagonal network. Beside being an interesting problem of its own, it will also give us insight into the mechanisms governing the formation of the dissolution patterns. Let us then set initially all the diameters and wormhole lengths in the network to uniform values, $d=d_0$ and  $l=l_0$ throughout the whole system. Then, to induce a localized growth of such a system, we make a single small cut in the inlet region, increasing the diameter by a factor of four  in a vertical column comprising 10 network nodes situated at the center of the inlet manifold (cf. Fig.~\ref{fig:symmetric}). Fig.~\ref{fig:symmetric} presents the dissolution patterns obtained for $G=1$ and different $\text{Da}_{eff}$ in such a system (for the simulation without pore merging). Two strikingly different patterns can be observed there: for small $\text{Da}_{eff}$ a  dendrite-like wormhole is formed, with a characteristic, spearhead-like shape. In this case the width of the wormhole is much larger than the pore scale.  Around $\text{Da}_{eff}^\star=0.54$, a sudden transition takes place, to an 'inverted Y' structure,  involving two channels growing sideways at equal angles from the main branch. The channels are only one pore wide and almost all the dissolution and flow becomes quickly concentrated there.  The value of $G$ affects the patterns to a weak extent only, its main effect being to shift the transition point between the patterns: for $G=10$ the critical $Da_{eff}$ value becomes $\text{Da}_{eff}^\star=0.67$, whereas for $G=0.1$ it is $\text{Da}_{eff}^\star=0.35$. This is in agreement with the observations discussed in Sec.~\ref{phase_diag}: larger $G$ values limit the dissolution at the inlet and thus increase the reactant penetration length and allow for the formation of more diffuse dissolution patterns.

\begin{figure}
\centering
\includegraphics[angle=0,width=5cm]{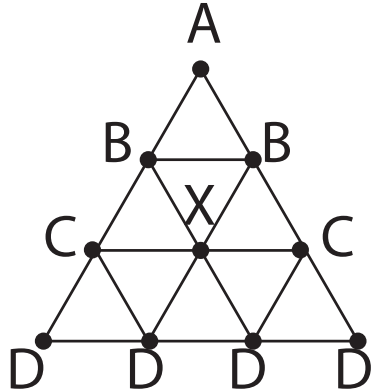}
\caption{A simplified triangular network with a single input node (A) and line output (D).}
\label{fig:schemat_sym}
\end{figure}

To elucidate the origin of the patterns observed in the dissolution of regular lattices, let us  consider a simplified triangular network shown in Fig.~\ref{fig:schemat_sym}. Here A represents the tip of the cut, with $p_A=1$ whereas $D$ represents the outlet boundary of the system, with $p_D=0$. By solving the Kirchoff equations in this case, one gets $\frac{P_B-P_C}{P_B-P_X}=\frac{7}{6}$. The flow along side pores is therefore slightly larger than in the central ones at the beginning of the simulation, when resistances of all the pores are the same. However, at large $\text{Da}_{eff}$, when almost all the reactant is spent in the first pores,  even a small difference in the flows will be enhanced by the dissolution. As a result, the hydrodynamic resistance of a channel BC will decrease much faster than that of BX, and the concentration of reactant at $C$ will be significantly larger than that in $X$. Thus the line $AD$ will continue to grow and soon all of the flow in the system will be concentrated there. 
On the other hand, for smaller $\text{Da}_{eff}$ both BX and BC will dissolve almost uniformly. In that case, the concentration of reactant in $X$ will be in fact larger than that in $C$ (since $X$ is being fed by two reactant-bearing pores instead of one). The dissolution will then be concentrated in the central channels, and a dendrite-shaped wormhole will be formed.

We expect the shape of the dendrite in this case to be largely independent of the underlying lattice, and in fact similar shapes are observed in the case of random lattices in Fig.~\ref{fig6}. For a regular lattice, an analytical derivation of such a shape should in principle be possible,  which remains a future
task. On the other hand, the inverted Y pattern formed at large $\text{Da}_{eff}$ is strongly connected with the underlying network. The arms  of the 'inverted Y' figure propagate along the lattice directions and the pattern will look differently if the network is changed. It is also strongly unstable, as the arms begin to compete with each other, as can be observed in  Fig.~\ref{fig:symmetric}.

The above considerations, although originally pertaining to the regular network, have, in fact, a more general relevance, since they illustrate two basic mechanisms that govern the emergence of dissolution patterns in the network. At small or intermediate $\text{Da}_{eff}$ diffuse, multi-pore patterns are formed in the extended region where the initial concentration of reactant is large. For large effective Damk\"ohler numbers the local pressure drop effects become more important and the dissolution is focused along a thin pore-wide wormhole which is more irregular and branched, reflecting the disorder of initial network.

\section{Optimal injection rate}\label{breakthrough}

\begin{figure}[!t,po_calosci_ewolucja]
\centering
\includegraphics[angle=0,width=6cm]{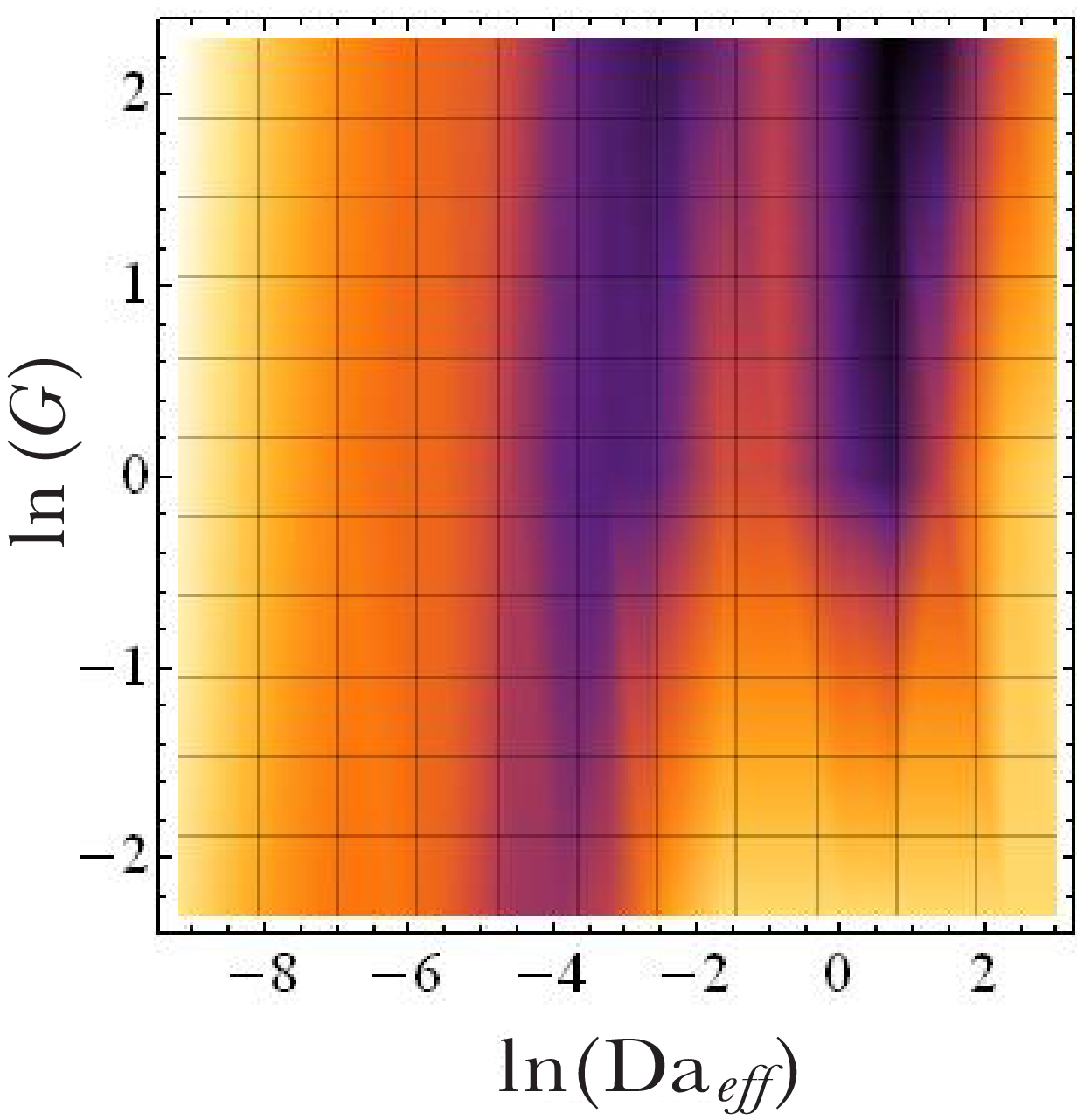}
\includegraphics[angle=0,width=1.5cm]{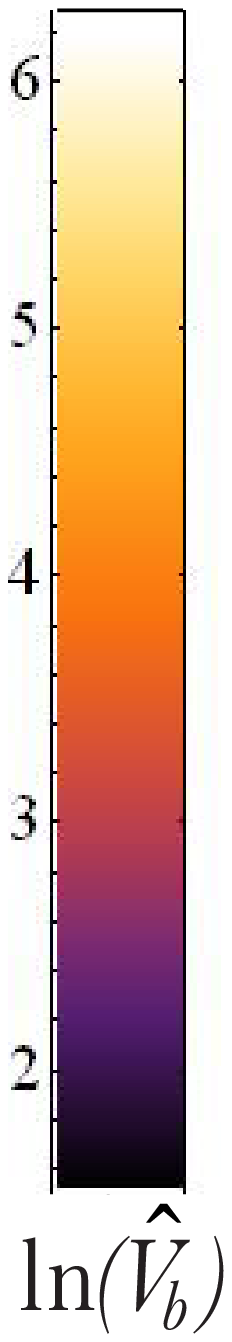}
\caption{
(Color online) Pore volume to breakthrough for different $\text{Da}_{eff}$ and $G$ for the simulation with fixed total flow on 200 $\times$ 200 lattice (broadening factor $\beta = 4$).}
\label{fig:czas_przestrzennie_Q}
\end{figure}

Understanding of the emergence of the dissolution patterns is particularly important in optimization of carbonate reservoir stimulation treatments, where the relevant question is how to get the maximum increase of permeability for a given amount of reactive fluid. Numerical and experimental investigations of reactive flows in porous media~\cite{Fredd1998,Golfier2002b,Panga2005,Kalia2007,Cohen2008a} suggest
that there exists an optimum injection rate, which maximizes the permeability gain for a given amount of reactant. If the injection rate is relatively small, a large portion of the reactant is wasted by the unproductive dissolution of the inlet pores while the overall increase in permeability remains small. On the other
hand, for very large injection rates, the reactant is exhausted on a uniform opening of all of the pores in the system, which is also inefficient
in terms of permeability increase. The optimum flow rate must give rise to spontaneous channeling, since the reactant is then used to create a small number of highly permeable wormholes, which transport the flow most efficiently. To
quantify the optimization with respect to $\text{Da}_{eff}$ and $G$, we measured the total volume of reactive fluid, $V_{b}$, that must be injected into the pores to obtain the breakthrough, which we identify with the moment when the diameter of at least one of the outlet pores increases by a factor of $\beta$. It is convenient to express 
 $V_{b}$ in terms of a dimensionless quantity
$$
\hat{V}_b = \frac{\gamma V_b}{V_0} =  \frac{\gamma T_b Q}{V_0},
$$
where $\gamma$ is the acid capacity number \eqref{acid}, $V_0$ is the initial total fluid volume in the network, $Q$ is a (constant) injection rate and $T_b$ is the breakthrough time. A diagram of $\hat{V}_b$ as a function of $\text{Da}_{eff}$ and $G$ is presented in Figure \ref{fig:czas_przestrzennie_Q}. In these simulations we did not allow for the pore merging. 

As elucidated above, for both small and large effective Damk\"{o}hler numbers $\hat{V}_b$ is relatively large. However, the value of $G$ also has a non-trivial impact on $\hat{V}_b$. Namely, for small values of $G$ (in reaction-limited regime) the dissolution in the inlet region is strongly enhanced, particularly for large Damk\"{o}hler numbers, which leads to the high consumption of reactant there and the associated increase in $\hat{V}_b$. 

\begin{figure}[!t,po_calosci_ewolucja]
\centering
\includegraphics[angle=0,width=8.5cm]{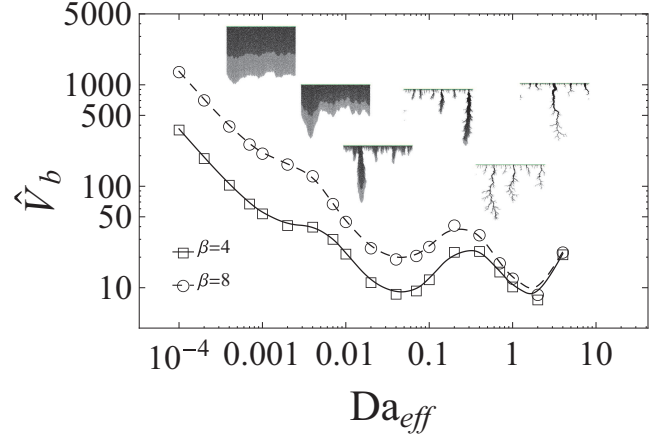}
\caption{
Breakthrough volume as a function of $\text{Da}_{eff}$ for $G=1$ for the simulation with fixed total flow on 200 $\times$ 200 lattice.
The data for two different conditions for breakthrough is presented: $\beta = 8$ (circles) and $\beta = 4$ (squares). The lines are a guide to the eye.}
\label{fig:czas_podsumowanie}
\end{figure}

In the real physical system, the value of $G$ cannot be varied during the experiment, since both the diffusion constant and reaction rate are material properties. Thus, changing the injection rate moves the system along a line of constant $G$, sweeping the range of $\text{Da}_{eff}$ values. This leads to $\hat{V}_b(\text{Da}_{eff})$ dependence, the example of which, for $G=1$, is presented in Fig.~\ref{fig:czas_podsumowanie}. Two different conditions for breakthrough are compared here, corresponding to  $\beta = 4$ and $\beta = 8$, again without  pore merging. A stronger breakthrough condition results in an increase of $\hat{V}_b$, especially for smaller $\text{Da}_{eff}$. For larger $Da_{eff}$, however, a pronounced wormhole is formed in the system and at the moment of its breakthrough the dissolution in the outlet pores becomes so dynamic that $\hat{V}_b$ becomes almost independent of $\beta$.

The most striking feature of Fig.~\ref{fig:czas_podsumowanie} is the presence of  multiple minima on $\hat{V}_b(\text{Da}_{eff})$ curve. To interpret this finding, let us first analyse the dissolution of a single, long channel of length $L$ spanning the whole length of the system and comprised of $N_x=L/l_0$ elementary pores. As derived in the Appendix, if initially the channel is uniform, $d(x,t=0)=d_0$ and the concentration at the inlet is given by $c(x=0,t)=c_{in}$, then the diameter of the channel $d(x,t)$ is given implicitly by~\eqref{impl}
\begin{equation}
(1+G) \text{Da}^{chann}_{eff} \frac{x}{L} + G  \ln \left(\frac{\hat{d}^2-1}{\hat{d}_{in}^2-1} \right) -\\ 2 \left( \text{arccoth}(\hat{d})- \text{arccoth} (\hat{d}_{in}(\hat{t})) \right) = 0,
\label{eq:d_analitycznie}
\end{equation}
where $\hat{d}=d/d_0$ is the diameter of the channel scaled by the initial diameter $d_0$ and $d_{in}(\hat{t})$ is the diameter at the inlet $(x=0)$ which evolves according to~\eqref{din}
\begin{equation}
\hat{d}_{in} =   \frac{\sqrt{2G\hat{t}+G (G+2)+1}-1}{G}.
\label{din0}
\end{equation}
Finally, the effective Damk\"ohler number for the channel is given by the formula analogous to \eqref{Daeff}, i.e.
\begin{equation}
\text{Da}^{chann}_{eff}=\frac{\pi d_0 k L / q_{chann}}{1+G}
\label{eq:Dac}
\end{equation} 

\begin{figure}[!t,po_calosci_ewolucja]
\centering
\includegraphics[angle=0,width=8.5cm]{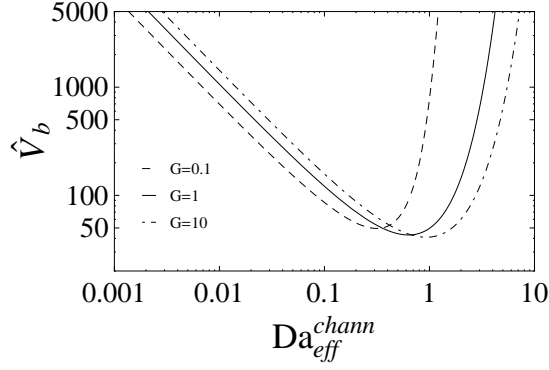}
\caption{
Pore volume to breakthrough as a function of $\text{Da}^{chann}_{eff}$ for $G=0.1$ (reaction-limited case, dashed  line) $G=1$ (mixed case, solid) and $G=10$ (diffusion-limited case, dot-dashed) based on the analytical solution for the profile of a single dissolving channel~\eqref{eq:d_analitycznie}. The value of the broadening factor used here is $\beta =4$.}
\label{fig:analitycznie_poj_rura}
\end{figure}

Defining breakthrough time $T_b$ as before as the moment at which the end of a channel $(x=L)$ has broadened $\beta$ times, we obtain  $\hat{V}_b(\text{Da}^{chann}_{eff})$ curves presented in Figure \ref{fig:analitycznie_poj_rura} for different $G$ regimes ($G= 0.1, 1$ and $10$). Interestingly, even on a level of a single tube we observe the minimum in  $\hat{V}_b(\text{Da}^{chann}_{eff})$ dependence. The increase of $\hat{V}_b$ at large $\text{Da}^{chann}_{eff}$ is caused by the nonuniform dissolution of the tube in the low-flow limit: the inlet dissolves faster than the outlet, which leads to the increase of the reactive surface area in the inlet region and further depletion of the reactant from the downstream regions. On the other hand, the increase of $\hat{V}_b(\text{Da}^{chann}_{eff})$ at very high flows (low $\text{Da}^{chann}_{eff}$) is connected with the fact that a significant portion of the reactant is then simply flushed through the system, not reacting with the walls. The minimum in  $\hat{V}_b(\text{Da}^{chann}_{eff})$ is present for all values of $G$, both in reactive-limited and transport-limited regime, though its position shifts to larger $\text{Da}_{eff}^{chann}$ as $G$ is increasing. 

It is interesting to note that although in general \eqref{eq:d_analitycznie} cannot be solved explicitly, for a number of specific values of $G$ a closed form expression for $T_b(\text{Da}_{eff})$ can be found. These include, among others, $G=0$ and $G=\infty$, i.e. reaction-limited and transport-limited case respectively, which are worked out in the Appendix, but also $G=1$, which corresponds to the case when both diffusion and reaction are important. Putting $x=L$ and $\hat{d}=\beta$ into \eqref{eq:d_analitycznie} we obtain in the latter case
\begin{equation}
\hat{T}_b = \frac{\beta-1}{2} e^{\text{Da}^{chan}_{eff}} \bigl( 4 + (\beta-1) e^{\text{Da}^{chann}_{eff}} \bigr)
\label{gabbel}
\end{equation}
or, in terms of the volume of reactive fluid injected, $V_b = q_{chann} T_b = q_{chann} d_0 \hat{T_b}/2 k \gamma$ 
\begin{equation}
\hat{V}_b = \frac{\gamma V_b}{V_0} = \frac{\beta-1}{2 \text{Da}^{chann}_{eff}} e^{\text{Da}^{chan}_{eff}} (4 + (\beta-1) e^{\text{Da}^{chann}_{eff}}),
\end{equation}
where we have normalized $V_b$ by the initial volume of the channel, $V_0 = \pi L d_0^2/4$. The above corresponds to the solid line in Fig.~\ref{fig:analitycznie_poj_rura}.

\begin{figure}[!t,po_calosci_ewolucja]
\centering
\includegraphics[angle=0,width=8.5cm]{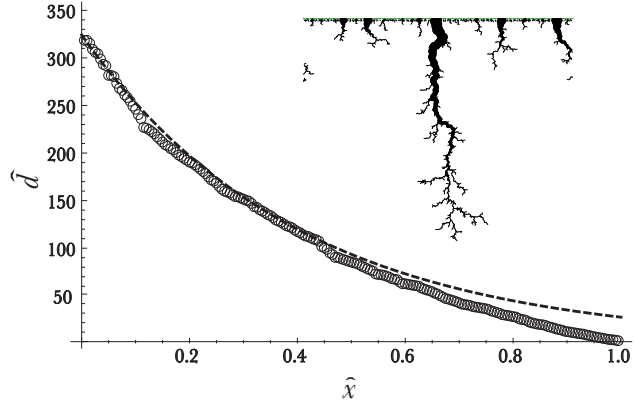}
\caption{
Dissolution of a channel obtained by numerical simulation on the network (empty circles) and exact analytical solution for single wormhole, Eq.~\eqref{eq:d_analitycznie} (dashed line). Comparison was made for $\text{Da}_{eff} = 5$ and $G=1$ at the breakthrough time.}
\label{fig:rura_dopasowanie}
\end{figure}

The single-channel model should adequately describe the results of the network dissolution in the regime of large $\text{Da}_{eff}$, where the competition between the flowpaths is strong and leads to the emergence of a solitary winning wormhole. 
Figure \ref{fig:rura_dopasowanie} shows the profile of a channel formed in such a regime (for $\text{Da}_{eff} = 5$ and $G=1$) at the breakthrough time. As observed, the profile of the channel agrees with the analytical result \eqref{eq:d_analitycznie} very well, except in the region of large $x$. This is caused by the fact that in the vicinity of channel tip there is an intensive leakoff from the channel towards the matrix \cite{Szymczak2006}, thus the assumption of the constant flow throughout the channel, used in derivation of \eqref{eq:d_analitycznie} ceases to be valid.

\begin{figure}[!t,po_calosci_ewolucja]
\centering
\includegraphics[angle=0,width=6cm]{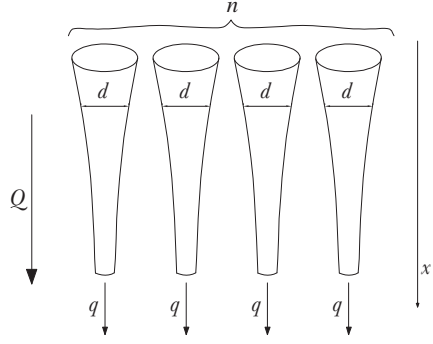}
\caption{
A schematic of a simple model of disconnected channels.}
\label{fig:model_pojedynczych_kanalow}
\end{figure}

At smaller Damk\"{o}hler numbers,  there are many alternative flowpaths between inlets and outlets which divide the available flow between themselves.  In the simplest approximation, we can assume that the flow is divided equally between the flowpaths ({\it cf.} Fig.~\ref{fig:model_pojedynczych_kanalow}) and then  again apply Eq.~\eqref{eq:d_analitycznie} to one of such flowpaths.  However, in order to calculate the Damk\"{o}hler number \eqref{eq:Dac} in one of these channels we need to assess what percentage of the total flow goes through a particular path. This depends on the amount of focusing in the system, which, as demonstrated in Sec. \ref{phase_diag}, is a function of the overall Damk\"{o}hler number, $\text{Da}_{eff}$, characterizing the whole system: for small $\text{Da}_{eff}$ the dissolution is uniform throughout the system and $q_{chann}$ will be equal to $Q/N_{inlet}$. For larger $\text{Da}_{eff}$, undulations begin to be formed on the dissolution front, but the emerging fingers are relatively thick,  comprised of many parallel pores. For even larger $\text{Da}_{eff}$  the fingers become narrower and finally, for the largest $\text{Da}_{eff}$, only one pore at each height carries all the fluid, and $q_{chann} = Q$.

\begin{figure}[!t,po_calosci_ewolucja]
\centering
\includegraphics[angle=0,width=8.5cm]{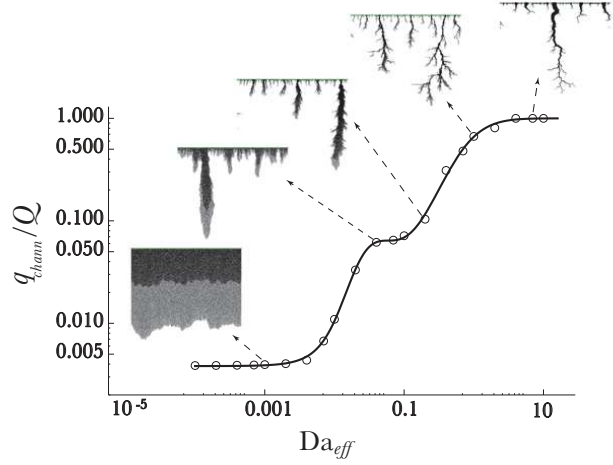}
\caption{The characteristic flow through an active flowpath, $q_{chann}$, as a function of $\text{Da}_{eff}$ together with a phenomenological fit, Eq.~\eqref{eq:fit}.
}
\label{fig:ile_kanalow_Da}
\end{figure}

In general, the percentage of the flow focused in the active flowpaths, $q_{chann}/Q$, is an increasing function of $\text{Da}_{eff}$ of a rather complicated nature. Fig.~\ref{fig:ile_kanalow_Da} presents an example of such a function numerically estimated for $G=1$ together with a phenomenological fit to the data:
\begin{equation}
\ln \left( q_{chann}/Q \right)=   \ln \left( N_{inlet} \right) \Big( \frac{1}{2} (\tanh (a_1 + \ln \text{Da}_{eff})-1)   \\ - a_2 e^{-\frac{(\ln \text{Da}_{eff}+a_3)^2} {a_4}}   \Big),
\label{eq:fit}
\end{equation}
with   $a_1=4.10$, $a_2 = 0.42$,  $a_3 = 2.34$, and $a_4 =3.03$.

An interesting feature of the dependence $q_{chann}(\text{Da}_{eff})$ is a plateau for intermediate values of $\text{Da}_{eff}$ (starting from $\text{Da}_{eff} \approx 0.02$).  The analysis of the patterns in this region shows that it corresponds to the onset of branching of the dissolution channels. However, both  the number of active channels and their widths remain approximately constant up to about $\text{Da}_{eff}\approx 0.2$.

Combining the $q_{chann}(\text{Da}_{eff})$ dependence of Eq.\eqref{eq:fit} with~\eqref{gabbel} leads to
\begin{equation}
\hat{T}_b = \frac{\beta-1}{2} e^{\frac{\text{Da}_{eff} N_x Q}{2 N_y q_{chann}}} \bigl(4 + (\beta-1) e^{\frac{\text{Da}_{eff} N_x Q}{2 N_y q_{chann}}} \bigr),
\label{gabbel2}
\end{equation}
where we have linked the channel Damk\"{o}hler number \eqref{eq:Dac} to the network Damk\"{o}hler number \eqref{eq:Da} by
\begin{equation}
Da_{eff}^{chann} = \frac{\pi k d_o L}{q_{chann}(1+G)} = \frac{\pi k d_o l_0}{q_0 (1+G)} \frac{N_x q_0}{q_{chann}} = 
 \frac{Da_{eff} N_x Q}{2 N_{y} q_{chann}}.
\end{equation}
In the above, we have used the fact that $L=l_0 N_x$ and $Q = N_{inlet} q_0 = 2 N_y q_0$, since there are two channels leading out of each inlet node ({\it cf.} Fig.~\ref{fig:schemat_po_calosci}). 
In terms of the injected volume (still for $G=1$)
\begin{equation}
\hat{V}_b =  \frac{\gamma V_b}{V_0} = \frac{\beta-1}{3 N_x \text{Da}_{eff}} e^{\frac{\text{Da}_{eff} N_x Q}{2 N_y q_{chann}}} \bigl( 4 + (\beta-1) e^{\frac{\text{Da}_{eff} N_x Q}{2 N_y q_{chann}}} \bigr),
\label{gabbel2b}
\end{equation}
where $\hat{V}_b$ is normalized by the initial volume of the pores in a hexagonal $N_x \times N_y$ lattice, $V_0 = 3N_x N_y \pi l_0 d_0^2/4$. Note that $q_{chann}$ in the above formulae is in itself a function of both $Da_{eff}$ and the system size, as given by Eq.~\eqref{eq:fit}. 

\begin{figure}[!h,przykladowa_siatka]
\centering
\includegraphics[angle=0,width=8.5cm]{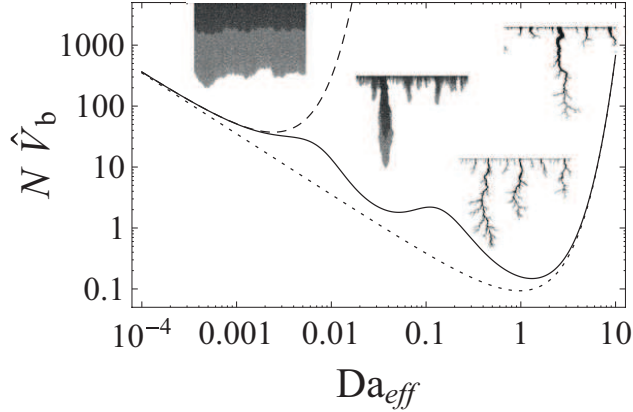}
\caption{ Pore volume to breakthrough obtained for a simple model of non-interacting channels, using Eq.~\eqref{eq:d_analitycznie} with the flow in the active channel, $q_{chann}$ obtained based on Eq.~\eqref{eq:fit}. The dotted line corresponds to \eqref{eq:d_analitycznie}  with $q_{chann}=Q$ (all flow focused in a single channel) whereas the dashed line corresponds to the uniform dissolution with $q_{chann}=Q/N_{inlet}$}
\label{fig:analitycznie_czasy_przebicia}
\end{figure}

The resulting $\hat{V}_b(Da_{eff})$ dependence for 200 $\times$ 200 system is presented in Fig.~\ref{fig:analitycznie_czasy_przebicia}. Additionally, we plot here the $\hat{V}_b(\text{Da}_{eff})$ dependences obtained under the assumption that the entire flow in the system is always focused in a single channel, irrespectively of the Damk\"{o}hler number, i.e.  $q_{chann}=Q$ and yet another curve corresponding to $q_{chann}=Q/N_{inlet}$, i.e. with the assumption that there is no flow focusing at all and the fluid flows through all available flowpaths.  The comparison of these dependences helps us to understand the origin of particular minima in $\hat{V}_b(\text{Da}_{eff})$  curve. The deepest minimum at $\text{Da}_{eff}\approx 1$ corresponds to the optimal condition of dissolution in a single channel model with $q_{chann}=Q$, i.e. to the case when the entire flow is focused in a single flowpath. On the other hand, the shoulder at the left, near $\text{Da}_{eff} \approx 0.002$ corresponds to the optimal conditions of dissolution in the case when all the flowpaths dissolve uniformly. Finally, the local minimum at $\text{Da}_{eff} \approx 0.03$ corresponds to the plateau region on $q_{chann}(\text{Da}_{eff})$ curve and is an effect of superimposing the dependences in Figs. \ref{fig:analitycznie_poj_rura} and \ref{fig:ile_kanalow_Da}. All of the above-discussed features are also visible in the $\hat{V}_b(\text{Da}_{eff})$ curves constructed from the numerical data in Fig.\ref{fig:czas_podsumowanie}.

\begin{figure}[!t,po_calosci_ewolucja]
\centering
\includegraphics[angle=0,width=8.5cm]{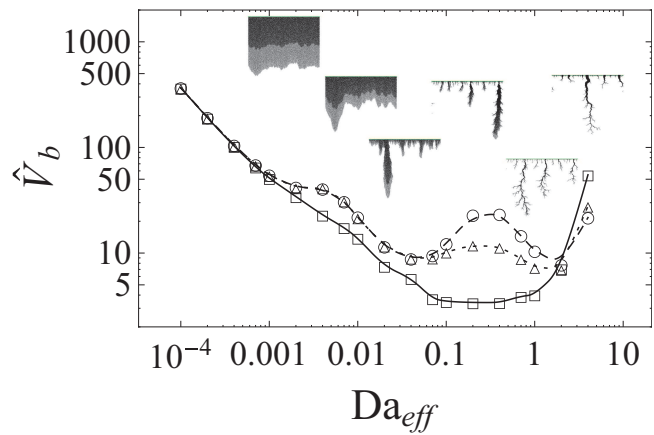}
\caption{
Breakthrough volume as a function of $\text{Da}_{eff}$ for $G=1$ and $\beta=4$ for the simulation without pore merging (circles) and with pore merging. In the latter case, two values of the initial pore aspect ratio are considered: $d_0/l_0=0.025$ (triangles) and $d_0/l_0 = 0.1$ (squares). The simulations were carried out on 200 $\times$ 200 lattice.}
\label{fig:czas_podsumowanie2}
\end{figure}

The data discussed so far has been obtained without pore merging. Since the larger channel grows faster than two smaller ones, merging will in general decrease the breakthrough times.  
Fig.~\ref{fig:czas_podsumowanie2} presents the breakthrough curves for the simulation with pore merging for two different values of initial aspect ratio, $d_0/l_0=0.1$ and $d_0/l_0=0.025$.  Comparing these curves with that for a non-merging case, one observes a significant speed-up of dissolution and decrease of breakthrough times in the intermediate range of $\text{Da}_{eff}$. For small Damk\"{o}hler numbers the impact of merging on the dynamics is much weaker, since the pores then remain thin and thus hardly merge. On the other hand, for large $\text{Da}_{eff}$, there emerges a single wormhole which eats up the neighboring ones. In the merging simulations the main wormhole is broadening faster than without merging, particularly near the inlet. As already mentioned, this leads to a large consumption of reactant there which slows down the advancement of the dissolution front and gives rise to the increase of breakthrough times.

The main difference between the breakthrough curves in merging and non-merging case is however a gradual disappearance of multiple minima structure. Already at $d_0/l_0 = 0.025$ the barrier between the two main minima becomes lower and at $d_0/l_0 = 0.1$ it vanishes altogether and a single minimum at $\text{Da}_{eff} \simeq 0.2$ is observed. A similar dependence of $\hat{V}_{b}$ on Damk\"ohler number has been reported previously~\cite{Fredd1998,Golfier2002b,Panga2005,Kalia2007,Cohen2008a}, both in the experiments and simulations of porous media dissolution. 

\begin{figure}[!t]
\centering
\includegraphics[angle=0,width=8.5cm]{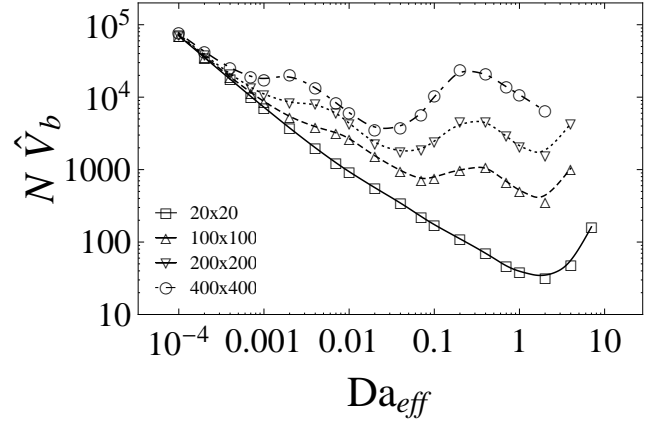}
\caption{
Breakthrough volume as a function of $ \text{Da}_{eff}$ for $G=1$ and $\beta=4$ for the simulation without pore merging for $N\times N$ lattice with $N=20$ (solid), $100$ (dashed), $200$ (dotted) and $400$ (dot-dashed). }
\label{size_nojoin}
\end{figure}

\begin{figure}[!t]
\centering
\includegraphics[angle=0,width=8.5cm]{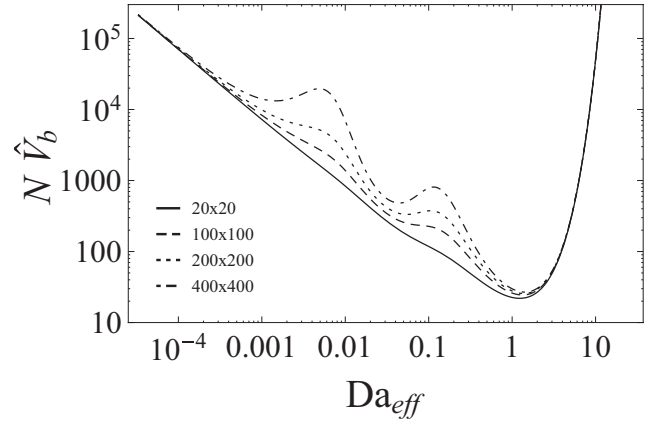}
\caption{
Pore volume to breakthrough obtained using a model of non-interacting channels (Eqs.~\eqref{eq:d_analitycznie}-\eqref{eq:fit} for $N\times N$ lattice with $N=20$ (solid), $100$ (dashed), $200$ (dotted) and $400$ (dot-dashed). }
\label{analit}
\end{figure}

Finally, we discuss the effects of the system size on the above results. Fig.~\ref{size_nojoin} shows the $\hat{V}_d(Da_{eff}) $ curves for the simulation without pore merging for various $N \times N$ lattices with $N$ ranging from $20$ to $400$. We find it convenient to plot $N \hat{V}_d$ as the dependent variable here, as then the curves asymptote to a single curve as $Da_{eff} \rightarrow 0$. As observed, for small system sizes only one minimum appears in $\hat{V}_d(Da_{eff})$ dependence. For larger sizes this minimum bifurcates into two minima, and then, as the size is further increased, a third shallow minimum appears for small $N Da_{eff}$ values.  This behavior can be further elucidated using the semi-analytical model of Eqs.~(\ref{eq:d_analitycznie}-\ref{gabbel2b}), as presented in Fig.~\ref{analit}. In particular, the  $\text{Da}_{eff} \rightarrow 0$ limit of \eqref{gabbel2} gives a small $Da_{eff}$ asymptote in the form
\begin{equation}
N \hat{V_b} \approx \frac{(\beta-1)\bigl(3+\beta \bigr)}{3 Da_{eff}},
\end{equation}
indeed independent of $N$. Moreover, as observed in Fig.~\ref{analit} in the large $Da_{eff}$ limit the curves also converge, this time to
\begin{equation}
N \hat{V_b} \approx \frac{(\beta-1)^2}{3 \text{Da}_{eff}} e^{\text{Da}_{eff}}.
\end{equation}
In the numerical simulations of Fig.~\ref{size_nojoin} this limit is hard to attain, since at the large Damk\"{o}hler numbers the dissolution is highly nonuniform which is a source of convergence problems in the linear solver. 
The analytical model shows a similar minimum structure to the numerical data - at small $N$ a single minimum is present, at $Da_{eff} \approx 1$, then the second one appears, at $Da_{eff} \approx 0.03$ and finally the third at $Da_{eff} \approx 0.001$. As before, the rightmost minimum corresponds to the case when the entire flow is focused in a single wormhole, the central minimum  corresponds to the plateau region on $q_{chann}(\text{Da}_{eff})$ and the leftmost minimum is connected with a uniform dissolution of the network. Importantly, the positions of these minima are almost entirely independent of the size of the system. 

It is worth to note, however, that the model of Eqs.(\ref{eq:fit}-\ref{gabbel2b}) gives only a qualitative agreement with the numerical data - even though the general three-minima structure is reproduced correctly, there are significant differences in relative depths of the minima and the heights of the barriers separating them. The main approximation in the model is the assumption that the percentage of the flow focused in the main channel \eqref{eq:fit} remains constant throughout the simulation. In reality, however, initially all the channels have similar flow rates, 
and then gradually the flow focuses in the few main channels. Thus the model of Eqs.~(\ref{eq:fit}-\ref{gabbel2b}) will in general overestimate the total dissolution rate in the main channels throughout the simulations, particularly for large Damk\"{o}hler numbers, where flow focusing is the strongest. This leads to much larger depth of the rightmost minimum in $\hat{V_b}(Da_{eff})$ than that observed in the simulations.

\begin{figure}[!t]
\centering
\includegraphics[angle=0,width=8.5cm]{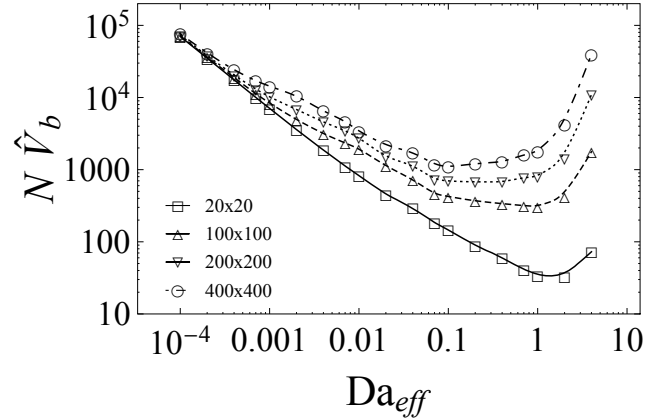}
\caption{
Breakthrough volume as a function of $\text{Da}_{eff}$ for $G=1$ and $\beta=4$ for the simulation with pore merging ($d_0/l_0=0.1$) for $N \times N$ lattice with $N=20$ (solid), $100$ (dashed), $200$ (dotted) and $400$ (dot-dashed).}
\label{size_join}
\end{figure}

Finally, let us discuss the size effects for the pore-merging simulations. The data presented in Fig.~\ref{size_join} shows that in this case there are only slight differences in $\hat{V}_b(Da_{eff})$ dependences as $N$ is increased. All of the curves have a single minimum only, which seems to be drifting towards smaller $Da_{eff}$ values as the size is increasing, however the minimum is rather wide so that in the entire region $0.1 \leq Da_{eff} \leq 1$ the values of breakthrough volumes are similar.

\section{Summary}

In this paper we have studied dissolution of porous medium using an evolving network model. The relevant dimensionless parameters to characterize dissolution patterns in this system are the effective Damk\"{o}hler number, $\text{Da}_{eff}$, relating the reaction rate to the mean fluid velocity in the pores and $G$, which measures the relative importance of diffusive and reactive phenomena.  As $\text{Da}_{eff}$ is increased, a transition from uniform dissolution to the strongly localized flow is observed, with the wormholes appearing at about $\text{Da}_{eff} \approx 0.02$. For intermediate values of $\text{Da}_{eff}$ (0.02-0.5) the wormholes are diffuse and involve many parallel pores. However, as $\text{Da}_{eff}$ is further increased, the wormhole diameter is drastically decreased and dissolution becomes focused along thin, pore-wide, highly branched flowpaths. This transition in wormhole shape is particularly dramatic and abrupt in the case of regular network, where a characteristic, 'inverse Y-shaped' channels appear at $\text{Da}_{eff} \approx0.5$ (for $G=1$). 

 We have also analyzed the problem of  finding an optimum flow rate that gives a maximum increase in permeability for a given amount of reactant. The dependence of $V_b$ (the volume of the fluid needed to open the system) on $\text{Da}_{eff}$ turned out to be nontrivial, with a number of local minima, corresponding to different regimes of the dissolution. Interestingly, if the pores are allowed to merge together as they dissolve, the $V_b(Da_{eff})$ curve becomes simpler, with a single minimum.

\appendix

\section{The growth of a single cylindrical channel}

Let us consider the problem of dissolution of a single, long channel of length $L$. The shape of the channel is described by its diameter $d(x,t)$ as a function of axial coordinate $x$ and time $t$. In principle, one can imagine such a channel as a collection of $N_x=L/l_0$ pores from Sec.~\ref{model} connected in a serial manner. Let us assume that initially the channel is uniform, $d(x,t=0)=d_0$ and the concentration at the inlet is given by $c(x=0,t)=c_{in}$.

The erosion equation, \eqref{eq:erosion} can then be written as
\begin{equation}
\partial_t \hat{d} =  \frac{1}{d_0}  \frac{2 k \gamma}{1+G \hat{d}} \hat{c}
\label{dh}
\end{equation}
where $\gamma={c_{in}}/{\nu c_{sol}}$, whereas $G=\frac{k d_0}{Sh D}$ and the dimensionless fields $\hat{c}=c/c_{in}$, $\hat{d}=d/d_0$  has been introduced. Next, the concentration balance. Eq.~\eqref{eq:balance} gives
\begin{equation}
\partial_{x} \hat{c} = - \frac{\pi d_0 k}{q} \frac{\hat{d}}{1+G \hat{d}} \hat{c}
\label{dc}
\end{equation}
Scaling time by $d_0/ 2 k \gamma$ and the axial coordinate by the channel length:
\begin{equation}
\hat{t} = \frac{2k \gamma t}{d_0}
\end{equation}
\begin{equation}
\hat{x} = \frac{x}{L}
\label{dcb}
\end{equation}
 leads to
\begin{equation}
\partial_{\hat{t}} \hat{d} =   \frac{1}{1+G \hat{d}} \hat{c}
\label{dh2}
\end{equation}
and
\begin{equation}
\partial_{\hat{x}} \hat{c} = - \frac{ \pi d_0 L k}{q} \frac{\hat{d}}{1+G \hat{d}} \hat{c} =\\ - (1+G) \text{Da}^{chann}_{eff} \frac{\hat{d}}{1+G \hat{d}} \hat{c},
\label{dc2}
\end{equation}
where  the effective Damk\"ohler number for the channel is given by
\begin{equation}
\text{Da}^{chann}_{eff}=\frac{\pi d_0 k L / q_{chann}}{1+G}
\label{eq:Dac2}
\end{equation} 
Differentiating (\ref{dh2}) with respect to $\hat{x}$ and using (\ref{dc2})  we get
\begin{equation}
\partial_{\hat{x}\hat{t}} \hat{d} = -(1+G) \text{Da}^{chann}_{eff}   \frac{\hat{d}}{(1+G \hat{d})^2} \hat{c} - \frac{G}{(1+G \hat{d})^2} \hat{c} \partial_{\hat{x}} \hat{d}
\end{equation}
Finally, expressing $\hat{c}$ in terms of $\partial_{\hat{t}} \hat{d}$ using again \eqref{dh2} leads to the equation in terms of $\hat{d}$ only
\begin{equation}
\partial_{\hat{x}\hat{t}} \hat{d} = - (1+G) \text{Da}^{chann}_{eff}   \frac{\hat{d}}{1+G \hat{d}} \partial_t \hat{d} - \frac{G}{1+G \hat{d}} \partial_{\hat{t}} \hat{d} \partial_{\hat{x}} \hat{d}
\end{equation}
or, regrouping the terms
\begin{equation}
\partial_{\hat{x}t} \left( \hat{d}(1+G \frac{\hat{d}}{2}) \right) + \partial_t \left(\frac{1}{2} (1+G) \text{Da}^{chann}_{eff}    \hat{d}^2 \right) =0
\end{equation}
Integrating over time leads to the relation
\begin{equation}
\partial_{\hat{x}} \left( \hat{d}(1+G \frac{\hat{d}}{2}) \right) +  \frac{1}{2} (1+G) \text{Da}^{chann}_{eff}   \hat{d}^2  =W(\hat{x})
\label{W1}
\end{equation}
The function $W(\hat{x})$ can be obtained from the initial condition: $\hat{d}(\hat{t}=0)=1$, which gives
\begin{equation}
W = \frac{1}{2} (1+G) \text{Da}^{chann}_{eff} 
\end{equation}
The solution of \eqref{W1} can only be obtained in an implicit form:
\begin{equation}
(1+G) \text{Da}^{chann}_{eff}  \hat{x} + G  \ln \left(\frac{\hat{d}^2-1}{d_{in}^2-1} \right) -\\  2 \left( \text{arccoth}(\hat{d}(\hat{x},\hat{t}))- \text{arccoth} (\hat{d}_{in}(\hat{t})) \right) = 0
\label{impl}
\end{equation}
where $\hat{d}_{in}(\hat{t})$ is the inlet diameter, which is the solution of \eqref{dh2} with $\hat{c} = 1$, i.e.
\begin{equation}
\partial_t \hat{d}_{in} =   \frac{1}{1+G \hat{d}_{in}}
\label{dh2a}
\end{equation}
which gives
\begin{equation}
\hat{d}_{in}(\hat{t}) =   \frac{\sqrt{2 G \hat{t}+G (G+2)+1}-1}{G}
\label{din}
\end{equation}

We shall now consider two limiting cases of  Eq.~\eqref{impl}: the reaction-limited case, $G \rightarrow 0$, and the transport limited case, In the reaction limit \eqref{impl} takes the form
\begin{equation}
\text{Da}^{chann}  \hat{x}  -  2  \left( \text{arccoth}(\hat{d})- \text{arccoth} (\hat{d}_{in}) \right) = 0
\end{equation}
where 
\begin{equation}
\text{Da}^{chann} = \lim_{G \rightarrow 0} \text{Da}^{chann}_{eff}  = \frac{\pi k L d_0}{q} 
\end{equation}
In this case \eqref{din} becomes simply
\begin{equation}
\hat{d}_{in}(\hat{t}) =   1+\hat{t}
\label{dinr}
\end{equation}
The above two equations can then be combined to yield
\begin{equation}
\hat{d}= \coth \left( \frac{1}{2} \text{Da}^{chann}  \hat{x}+ \text{arccoth} (1+\hat{t}) \right)
\label{pdfg}
\end{equation}
Interestingly, the above profile has a well-defined limit as $t \rightarrow \infty$, namely
\begin{equation}
\hat{d}_{\infty} = \coth \left( \frac{1}{2} \text{Da}^{chann} \hat{x} \right).
\label{limi}
\end{equation}
Thus, at each given point, the channel {\it ceases to grow} after a certain time. This reflects the fact that as the diameter of the channel increases, there is a growing reactive surface, and hence the concentration is consumed faster, particularly near the inlet. Thus, paradoxically, the concentration profile should recede as time goes on. The receding concentration profile can indeed be observed, since in the reaction-limited case
\begin{equation}
\hat{c} = \partial_t \hat{d} =\\ \left((\hat{t}+1) \sinh \left(\frac{1}{2} \text{Da}^{chann} \hat{x} \right)+\cosh \left(\frac{1}{4} \text{Da}^{chann} \hat{x} \right) \right)^{-2}
\end{equation}
which is a decreasing function of time. 

In this limit, we can also calculate analytically the breakthrough time, i.e. the time $\hat{T}_b$ such that $\hat{d}(\hat{x}=1,\hat{t}=\hat{T}_{b})=\beta$. Inverting Eq.~\eqref{pdfg} leads to  
\begin{equation}
\hat{T}_{b} = \coth \left(\text{arccoth}(\beta) -\frac{1}{2} \text{Da}^{chann}\right)-1
\end{equation}
which can be rewritten as
\begin{equation}
\hat{T}_{b} = \frac{(\beta-1) \left(1+ \coth \left(\frac{1}{2} \text{Da}^{chann}\right) \right)}{\coth \left(\frac{1}{2} \text{Da}^{chann}\right) - \beta}
\label{tre}
\end{equation}
The above formula can be applied only if the denominator is positive, which gives the condition
\begin{equation}
\text{Da}^{chann} < 2 \text{arccoth}(\beta)
\end{equation}
a direct consequence of an upper limit of diameter growth, as given by \eqref{limi}.

In the transport limited case, ${G \rightarrow \infty}$, Eq. \eqref{impl} takes the form
\begin{equation}
\hat{d} =  \sqrt{1+\frac{2\hat{t}}{G} e^{-\pi D \text{Sh} L \hat{x}/q}}
\end{equation}
where we have used the fact that $ \lim_{G \rightarrow \infty} \text{Da}^{chann}_{eff} = \pi D \text{Sh} L /q$. 
This time the diameters increase without a limit. The breakthrough time is then given by
\begin{equation}
T_{b}=\frac{1}{2} G \left(\beta^2-1\right) e^{\pi D \text{Sh} L/q}
\label{ttra}
\end{equation}
This time $\hat{T}_b$ is well-defined in the entire flow range, but it dramatically increases as $q \rightarrow 0$.

\acknowledgments

A.B. acknowledges the support of the Foundation for Polish Science (FNP)
through TEAM/2010-6/2 project co-financed by the EU European Regional Development Fund. P.S. acknowledges the support of the Polish Ministry of Science and
Higher Education (Grant No. N202023 32/0702). The photograph of the patterns obtained in the dissolution of plaster of Paris (Fig.~\ref{Ewers}) are courtesy of prof. Ralph Ewers.

%\bibliography{porousmedia}

\end{document}